\input harvmac.tex

\lref\Penrose{
R. Penrose, ``Nonlinear gravitons and curved twistor theory," Gen. Rel. Grav. 7, 31-52 (1976).
}

\lref\Sokatchev{
E. Sokatchev,
``A Superspace Action For N=2 Supergravity,''
Phys.\ Lett.\ {\bf B100}, 466 (1981).
}

\lref\HW{P. Horava and E. Witten, ``Heterotic and type I string dynamics from eleven dimensions," Nucl. Phys. {\bf B460} 506-524 (1996), hep-th/9510209.}

\lref\AHS{M. F. Atiyah, N. Hitchin and I. M. Singer, ``Self-duality in four-dimensional Riemannian geometry," Proc. Roy. Soc. London Ser. A {\bf 362}, 425-461 (1978).}

\lref\bhfive{A. Chou, R. Kallosh, J. Ramfeld, S. Rey, M. Shmakova and W.K. Wong, ``Critical points and phase transitions in 5d compactifications of M-theory,'' Nucl. Phys. {\bf B508} 147-180 (1997), hep-th/9704142.}

\lref\OV{H. Ooguri and C. Vafa, {``Geometry of N=2 strings,''} {Nucl. Phys.}  {\bf B361}, 469-518 (1991).} 

\lref\AKV{M. Aganagic, A. Klemm and C. Vafa,
{``Disk instantons, mirror symmetry and the duality web,''}
Z. Naturforsch.  {\bf A57}, 1-28 (2002), hep-th/0105045.}

\lref\HM{J.~A.~Harvey and G.~W.~Moore,
``Superpotentials and membrane instantons,'' hep-th/9907026.}

\lref\DVV{R. Dijkgraaf, E. Verlinde and M. Vonk,
{``{O}n the partition sum of the {NS} five-brane,''} hep-th/0205281.} 

\lref\torrea{C. Torre, {``Perturbations of gravitational instantons,''}
{Phys. Rev.}  {\bf D41}, 3620 (1990).}

\lref\torreb{C. Torre, {``A topological field theory of gravitational
instantons,''}
{Phys. Lett.}  {\bf B252}, 242-246 (1990).}
   
\lref\AV{
M.~Aganagic and C.~Vafa,
``G(2) manifolds, mirror symmetry and geometric engineering,''
hep-th/0110171.}
   
\lref\ANU{M. Abe, A. Nakamichi and T. Ueno, {``Moduli space of topological
two form gravity,''}
{Mod. Phys. Lett.} {\bf A9}, 895-901 (1994), {hep-th/9306130}.} 

\lref\LNU{H. Y. Lee, A. Nakamichi and T. Ueno, {``Topological two form
gravity in four dimensions,''}
{Phys. Rev.} {\bf D47}, 1563-1568 (1993), {hep-th/9205066}.} 

\lref\Urbantke{H. Urbantke, {J. Math. Phys.} {\bf 25}, 2321 (1983).}

\lref\Israel{W. Israel, {\it Differential Forms in General Relativity,
Comm. Dublin Inst. Adv. Stud. Series A} {\bf 26} (Dublin).}

\lref\Horowitz{G. Horowitz, {``Exactly soluble diffeomorphism invariant
theories,''}
{Commun. Math. Phys.}  {\bf 125}, 417 (1989).} 

\lref\Plebanski{M. Plebanski, {``On the separation of {E}insteinian
substructures,''} {J. Math. Phys.}  {\bf 18}, 2511 (1977).} 

\lref\CDJM{R. Capovilla, T. Jacobson, J. Dell and L. Mason, {``Self-dual 2-forms and gravity,''}
Class. Quant. Grav. {\bf 8}, 41-57 (1991).}

\lref\WB{N. Berkovits and E. Witten,
{``Conformal supergravity in twistor-string theory,''} hep-th/0406051}. 

\lref\attrtwo{S. Ferrara, R. Kallosh and A. Strominger,
``{N=2} extremal black holes,'' Phys. Rev.  {\bf D52}, 5412-5416 (1995), hep-th/9508072.} 

\lref\attrone{A. Strominger, {``{M}acroscopic {E}ntropy
of {N=2} {E}xtremal {B}lack {H}oles,''} Phys. Lett.  {\bf B383}, 39-43 (1996), hep-th/9602111.} 

\lref\Kalloshb{S.~Ferrara and R.~Kallosh,
``Universality of Supersymmetric Attractors,''
Phys.\ Rev.\ D {\bf 54} (1996) 1525, hep-th/9603090.}

\lref\Kallosha{S.~Ferrara and R.~Kallosh,
``Supersymmetry and Attractors,''
Phys.\ Rev.\ D {\bf 54} (1996) 1514, hep-th/9602136.}

\lref\Wittentwistor{E.~Witten,
``Perturbative gauge theory as a string theory in twistor space,''
hep-th/0312171.}

\lref\CSiegel{G.~Chalmers and W.~Siegel,
``The self-dual sector of {QCD} amplitudes,''
Phys.\ Rev.\ {\bf D54}, 7628 (1996), hep-th/9606061.}

\lref\Siegel{W.~Siegel,
``N=2, N=4 string theory is selfdual N=4 Yang-Mills theory,''
Phys.\ Rev.\ {\bf D46}, 3235 (1992);
``Selfdual N=8 supergravity as closed N=2 (N=4) strings,''
Phys.\ Rev.\ {\bf D47}, 2504 (1993), hep-th/9207043.}

\lref\GopakumarV{R.~Gopakumar and C.~Vafa,
``Topological gravity as large N topological gauge theory,''
Adv.\ Theor.\ Math.\ Phys.\  {\bf 2}, 413 (1998), hep-th/9802016;
``On the gauge theory/geometry correspondence,''
Adv.\ Theor.\ Math.\ Phys.\  {\bf 3}, 1415 (1999), hep-th/9811131.}

\lref\GVtwo{R. Gopakumar and C. Vafa, "M-theory and topological strings. II," hep-th/9812127.}

\lref\ADKMV{M. Aganagic, R. Dijkgraaf, A. Klemm, M. Marino and C. Vafa,
{``{T}opological strings and integrable hierarchies,''} {hep-th/0312085}.}

\lref\AAMV{M. Aganagic, A. Klemm, M. Marino and C. Vafa,
"The topological vertex," hep-th/0305132.}

\lref\VafaYM{C. Vafa, {``{T}wo dimensional {Y}ang-{M}ills,
black holes and topological strings,''} hep-th/0406058.}

\lref\CGLP{M. Cvetic, G. W. Gibbons, H. Lu and C. N. Pope, 
{``Bianchi IX self-dual Einstein metrics and singular G(2) manifolds,''}
Class. Quant. Grav.  {\bf 20}, 4239-4268 (2003), hep-th/0206151.}

\lref\AT{A.~Achucarro and P.~K.~Townsend, ``A Chern-Simons Action
For Three-Dimensional Anti-De Sitter Supergravity Theories,''
Phys.\ Lett.\ {\bf B180}, 89 (1986).}

\lref\Wittengrav{E.~Witten, ``(2+1)-Dimensional Gravity As An
Exactly Soluble System,'' Nucl.\ Phys.\ {\bf B311}, 46 (1988).}

\lref\OSV{H.~Ooguri, A.~Strominger and C.~Vafa, ``Black hole attractors and
the topological string,'' hep-th/0405146.}

\lref\ftheory{C.~Vafa, ``Evidence for F-Theory,''
Nucl. Phys. {\bf B469}, 403 (1996).}

\lref\Wittenhol{E.~Witten,
``Quantum Background Independence In String Theory,'' hep-th/9306122.}

\lref\Wittencs{E.~Witten,
``Chern-Simons gauge theory as a string theory,''
Prog.\ Math.\  {\bf 133}, 637 (1995), hep-th/9207094.}

\lref\WittenJones{
E.~Witten, ``Quantum Field Theory And The Jones Polynomial,''
Commun.\ Math.\ Phys.\  {\bf 121}, 351 (1989).}

\lref\BSadov{M.~Bershadsky and V.~Sadov,
``Theory of Kahler gravity,''
Int.\ J.\ Mod.\ Phys.\ A {\bf 11}, 4689 (1996), hep-th/9410011.}

\lref\TV{V.G.~Turaev and O.Yu.~Viro, ``State-sum invariants of
3-manifolds and quantum 6j-symbols,'' Topology {\bf 31}, 865 (1992).}

\lref\Turaev{V.G.~Turaev, C. R. Acad. Sci. Paris, t. 313,
S\'erie I 395 (1991); ``Topology of Shadow,'' preprint (1991).}

\lref\Walker{K.~Walker, ``On Witten's 3-Manifold Invariants,'' unpublished.}

\lref\RT{N.Y.~Reshetikhin and V.G.~Turaev,
``Invariants of 3-manifolds via link polynomials and quantum groups,''
Invent. Math. {\bf 103}, 547 (1991).}

\lref\Apol{S.~Gukov, ``Three-dimensional quantum gravity,
Chern-Simons theory, and the  A-polynomial,'' hep-th/0306165.}

\lref\AKMV{M.~Aganagic, A.~Klemm, M.~Marino and C.~Vafa,
``The topological vertex,'' hep-th/0305132.}

\lref\BCOVa{M.~Bershadsky, S.~Cecotti, H.~Ooguri and C.~Vafa,
``Holomorphic anomalies in topological field theories,''
Nucl.\ Phys.\ {\bf B405}, 279 (1993).}

\lref\BCOV{M.~Bershadsky, S.~Cecotti, H.~Ooguri and C.~Vafa,
``Kodaira-Spencer theory of gravity
and exact results for quantum string amplitudes,''
Commun.\ Math.\ Phys.\  {\bf 165}, 311 (1994).}

\lref\ORV{A.~Okounkov, N.~Reshetikhin and C.~Vafa,
``Quantum Calabi-Yau and classical crystals,'' hep-th/0309208.}

\lref\NV{A.~Neitzke and C.~Vafa, ``N=2 strings and the twistorial Calabi-Yau,'' hep-th/0402128.}

\lref\NOV{N.~Nekrasov, H.~Ooguri and C.~Vafa,
``S-duality and Topological Strings,'' hep-th/0403167.}

\lref\Hit{N.~Hitchin,
``The geometry of three-forms in six and seven dimensions,'' math.\-DG/0010054.}

\lref\Hitchin{N.~Hitchin, ``Stable forms and special metrics,'' math.DG/0107101.}

\lref\Mayr{P.~Mayr,
``$N=1$ Mirror Symmetry and Open/Closed String Duality,''
Adv.Theor.Math.Phys. {\bf 5}, 213 (2002).}

\lref\DT{S.K.~Donaldson and R.P.~Thomas,
``Gauge theory in higher dimensions,'' in {\it The Geometric Universe;
Science, Geometry, And The Work Of Roger Penrose}, Oxford University Press, 1998.}

\lref\MNOP{D.~Maulik, N.~Nekrasov, A.~Okounkov and R.~Pandharipande,
``Gromov-Witten theory and Donaldson-Thomas theory, I,'' math.AG/0312059.}

\lref\BLN{L.~Baulieu, A.~Losev and N.~Nekrasov,
``Chern-Simons and Twisted Supersymmetry in Higher Dimensions,''
Nucl. Phys. {\bf B522}, 82 (1998), hep-th/9707174.}

\lref\BKSi{L.~Baulieu, H.~Kanno and I.M.~Singer,
``Special Quantum Field Theories In Eight And Other Dimensions.''
Commun. Math. Phys. {\bf 194}, 149 (1998), hep-th/9704167.}

\lref\BKSii{L.~Baulieu, H.~Kanno and I.M.~Singer,
``Cohomological Yang-Mills Theory in Eight Dimensions,''
hep-th/9705127.}

\lref\Mms{A.~Losev, G.~Moore and S.~Shatashvili,
``M\&m's,'' Nucl. Phys. {\bf B522}, 105 (1998), hep-th/9707250.}

\lref\BT{M.~Blau and G.~Thompson,
``Aspects of $N_{T}\geq 2$ Topological Gauge Theories and D-Branes,''
Nucl. Phys. {\bf B492}, 545 (1997), hep-th/9612143.}

\lref\ALS{B.S.~Acharya, M.~O'Loughlin and B.~Spence,
``Higher Dimensional Analogues of Donaldson-Witten Theory,''
Nucl. Phys. {\bf B503}, 657 (1997), hep-th/9705138.}

\lref\Kovalev{A.~Kovalev,
``Twisted connected sums and special Riemannian holonomy,''
math.dg/0012189.}

\lref\BGGG{A.~Brandhuber, J.~Gomis, S.~S.~Gubser and S.~Gukov,
``Gauge theory at large N and new G(2) holonomy metrics,''
Nucl.\ Phys.\ {\bf B611}, 179 (2001), hep-th/0106034.}

\lref\Brandhuber{A.~Brandhuber,
``G(2) holonomy spaces from invariant three-forms,''
Nucl.\ Phys.\ {\bf B629}, 393 (2002), hep-th/0112113.}

\lref\GYZ{S.~Gukov, S.~T.~Yau and E.~Zaslow,
``Duality and fibrations on G(2) manifolds,'' hep-th/0203217.}

\lref\CCGLPW{
Z.~W.~Chong, M.~Cvetic, G.~W.~Gibbons, H.~Lu, C.~N.~Pope and P.~Wagner,
``General metrics of G(2) holonomy and contraction limits,''
Nucl.\ Phys.\ {\bf B638}, 459 (2002), hep-th/0204064.}

\lref\SYZ{A.~Strominger, S.~T.~Yau and E.~Zaslow,
``Mirror symmetry is T-duality,''
Nucl.\ Phys.\ {\bf B479}, 243 (1996), hep-th/9606040.}

\lref\Verlinde{E.~Verlinde,
``Global aspects of electric - magnetic duality,''
Nucl.\ Phys.\ {\bf B455}, 211 (1995), hep-th/9506011.}

\lref\Wab{E.~Witten,
``On S duality in Abelian gauge theory,''
Selecta Math.\  {\bf 1}, 383 (1995), hep-th/9505186.}

\lref\Wsltwoz{E.~Witten,
``SL(2,Z) action on three-dimensional conformal field theories
with Abelian symmetry,'' hep-th/0307041.}

\lref\BKS{L.~Baulieu, H.~Kanno and I.~M.~Singer,
``Special quantum field theories in eight and other dimensions,''
Commun.\ Math.\ Phys.\  {\bf 194}, 149 (1998) hep-th/9704167.}

\lref\BaulieuS{L.~Baulieu and I.~M.~Singer,
``Topological Yang-Mills Symmetry,'' Nucl.\ Phys.\ Proc.\ Suppl.\  {\bf 5B}, 12 (1988).}

\lref\Acharya{B.~S.~Acharya, M.~O'Loughlin and B.~Spence,
``Higher-dimensional analogues of Donaldson-Witten theory,''
Nucl.\ Phys.\ {\bf B503}, 657 (1997), hep-th/9705138.}

\lref\HPark{C.~Hofman and J.~S.~Park,
``Cohomological Yang-Mills theories on K\"ahler 3-folds,''
Nucl.\ Phys.\ {\bf B600}, 133 (2001), hep-th/0010103.}

\lref\BlauT{M.~Blau and G.~Thompson,
``Euclidean SYM theories by time reduction and special holonomy manifolds,''
Phys.\ Lett.\ {\bf B415}, 242 (1997), hep-th/9706225.}

\lref\VW{C.~Vafa and E.~Witten, ``A Strong coupling test of S duality,''
Nucl.\ Phys.\ {\bf B431}, 3 (1994), hep-th/9408074.}

\lref\Smolin{L.~Smolin,
``An invitation to loop quantum gravity,'' hep-th/0408048.}

\lref\MSpence{P.~de Medeiros and B.~Spence,
``Four-dimensional topological Einstein-Maxwell gravity,''
Class.\ Quant.\ Grav.\  {\bf 20}, 2075 (2003), hep-th/0209115.}

\lref\NakamichiI{A.~Nakamichi,
``Wave function of the universe in topological and in Einstein two form
gravity,'' hep-th/9303135.}

\lref\Perry{M.~J.~Perry and E.~Teo,
``Topological conformal gravity in four-dimensions,''
Nucl.\ Phys.\ {\bf B401}, 206 (1993) hep-th/9211063.}

\lref\Abe{M.~Abe, A.~Nakamichi and T.~Ueno,
``Gravitational instantons and moduli spaces in topological two form gravity,''
Phys.\ Rev.\ {\bf D50}, 7323 (1994) hep-th/9408178.}

\lref\joyce{D.~Joyce, ``Compact Manifolds with Special Holonomy'',
Oxford University Press, 2000.}

\lref\HL{R.~Harvey and H.B.~Lawson, Jr., ``Calibrated geometries",
Acta Math. {\bf 148}, 47 (1982).}

\lref\Mclean{ R.~Mclean, ``Deformations of Calibrated
Submanifolds'', Comm. Anal. Geom. {\bf 6}, 705-747 (1998).}

\lref\Bryant{R.L.~Bryant, ``Metrics with Exceptional Holonomy,''
Ann. Math. {\bf 126}, 525 (1987).}

\lref\BS{R.~Bryant and S.~Salamon,
"On the Construction of some Complete Metrics with Exceptional Holonomy",
Duke Math. {\bf J. 58}, 829 (1989).}

\lref\GPP{G. W.~Gibbons, D. N.~Page and C. N.~Pope, ``Einstein
Metrics on $S^3$, $R^3$ and $R^4$ Bundles,''
Commun. Math. Phys. {\bf 127}, 529-553 (1990).}

\lref\INOV{A.~Iqbal, N.~Nekrasov, A.~Okounkov and C.~Vafa,
``Quantum
foam and topological strings,'' hep-th/0312022.}

\lref\NOV{N.~Nekrasov, H.~Ooguri and C.~Vafa,
``S-duality and topological strings,'' hep-th/0403167.}

\lref\MNOP{ D. Maulik, N. Nekrasov, A. Okounkov and R. Pandharipande,
"Gromov-Witten theory and Donaldson-Thomas theory, I-II",
math.AG/0312059; math.AG/0406092.}

\lref\Vafa{C.~Vafa, ``Superstrings and Topological Strings at Large N,''
J. Math. Phys. {\bf 42}, 2798 (2001), hep-th/0008142.}

\lref\Louis{S.~Gurrieri, J.~Louis, A.~Micu and D.~Waldram,
``Mirror Symmetry in Generalized Calabi-Yau Compactifications,''
Nucl. Phys. {\bf B654}, 61 (2003), hep-th/0211102.}

\lref\CSalamon{S.~Chiossi and S.~Salamon, ``The intrinsic torsion of $SU(3)$ and $G_2$ structures,'' math.DG/0202282.}

\lref\Salamon{S.~Salamon, ``Almost Parallel Structures,''
Contemp. Math. {\bf 288}, 162 (2001), math.\-DG/0107146.}

\lref\Cardoso{
G.~L.~Cardoso, G.~Curio, G.~Dall'Agata, D.~Lust, P.~Manousselis and G.~Zoupanos,
``Non-K\"ahler string backgrounds and their five torsion classes,''
Nucl.\ Phys. {\bf B652}, 5 (2003), hep-th/0211118.}

\lref\Kachrunew{
S.~Kachru, M.~B.~Schulz, P.~K.~Tripathy and S.~P.~Trivedi,
``New supersymmetric string compactifications,''
JHEP {\bf 0303}, 061 (2003), hep-th/0211182.}

\lref\Gurrieri{S.~Gurrieri and A.~Micu,
``Type IIB theory on half-flat manifolds,''
Class.\ Quant.\ Grav.\  {\bf 20}, 2181 (2003), hep-th/0212278.}

\lref\Kaste{P.~Kaste, R.~Minasian, M.~Petrini and A.~Tomasiello,
``Nontrivial RR two-form field strength and $SU(3)$-structure,''
Fortsch.\ Phys.\  {\bf 51}, 764 (2003), hep-th/0301063.}

\lref\Becker{K.~Becker, M.~Becker, K.~Dasgupta and P.~S.~Green,
``Compactifications of heterotic theory on non-K\"ahler complex manifolds. I,''
JHEP {\bf 0304}, 007 (2003), hep-th/0301161.}

\lref\Gauntletttors{
J.~P.~Gauntlett, D.~Martelli and D.~Waldram, ``Superstrings with intrinsic torsion,''
Phys.\ Rev.\ {\bf D69}, 086002 (2004), hep-th/0302158.}

\lref\Fidanza{S.~Fidanza, R.~Minasian and A.~Tomasiello,
``Mirror symmetric $SU(3)$-structure manifolds with NS fluxes,''
hep-th/0311122.}

\lref\Grana{
M.~Grana, R.~Minasian, M.~Petrini and A.~Tomasiello,
``Supersymmetric backgrounds from generalized Calabi-Yau manifolds,''
JHEP {\bf 0408}, 046 (2004), hep-th/0406137.}

\lref\Gurrierii{S.~Gurrieri, A.~Lukas and A.~Micu,
``Heterotic on half-flat,'' hep-th/0408121.}

\lref\Kapustin{A.~Kapustin,
``Gauge theory, topological strings, and S-duality,''
JHEP {\bf 0409}, 034 (2004), hep-th/0404041.}

\lref\Giverntal{A.~Givental and B.~Kim,
``Quantum cohomology of flag manifolds and Toda lattices,''
Commun. Math. Phys. {\bf 168}, 609 (1995), hep-th/9312096.}

\lref\KManin{M.~ Kontsevich and Yu.~ Manin,
``Gromov-Witten classes, quantum cohomology, and enumerative geometry,''
Commun. Math. Phys. {\bf 164}, 525 (1994), hep-th/9402147.}

\lref\BManin{A.~Bayer and Yu.~Manin,
``(Semi)simple exercises in quantum cohomology,'' math.\-AG/0103164.}

\lref\Eguchietal{T.~Eguchi, K.~Hori and C.-S.~Xiong,
``Gravitational Quantum Cohomology,''
Int. J. Mod. Phys. {\bf A12}, 1743 (1997), hep-th/9605225.}

\lref\SGerasimov{A.~A.~Gerasimov and S.~L.~Shatashvili,
``Towards integrability of topological strings. I:
Three-forms on Calabi-Yau manifolds,'' hep-th/0409238.}

\lref\Hittwistor{N.J.~Hitchin, ``K\"ahlerian twistor spaces,''
Proc. London Math. Soc. {\bf 43}, no. 1, 133 (1981).}

\lref\Lebrun{C.~LeBrun, ``Twistors for tourists: a pocket
guide for algebraic geometers,''
Proc. Sympos. Pure Math., {\bf 62}, Part 2, 361 (1997).}

\def\boxit#1{\vbox{\hrule\hbox{\vrule\kern8pt
\vbox{\hbox{\kern8pt}\hbox{\vbox{#1}}\hbox{\kern8pt}}
\kern8pt\vrule}\hrule}}
\def\mathboxit#1{\vbox{\hrule\hbox{\vrule\kern8pt\vbox{\kern8pt
\hbox{$\displaystyle #1$}\kern8pt}\kern8pt\vrule}\hrule}}


\let\includefigures=\iftrue
\newfam\black
\includefigures
\input epsf
\def\figin{\epsfcheck\figin}\def\figins{\epsfcheck\figins}
\def\epsfcheck{\ifx\epsfbox\UnDeFiNeD
\message{(NO epsf.tex, FIGURES WILL BE IGNORED)}
\gdef\figin##1{\vskip2in}\gdef\figins##1{\hskip.5in}
\else\message{(FIGURES WILL BE INCLUDED)}%
\gdef\figin##1{##1}\gdef\figins##1{##1}\fi}
\def\DefWarn#1{}
\def\figinsert{\goodbreak\midinsert}
\def\ifig#1#2#3{\DefWarn#1\xdef#1{fig.~\the\figno}
\writedef{#1\leftbracket fig.\noexpand~\the\figno}%
\figinsert\figin{\centerline{#3}}\medskip\centerline{\vbox{\baselineskip12pt
\advance\hsize by -1truein\noindent\footnotefont{\bf Fig.~\the\figno:} #2}}
\bigskip\endinsert\global\advance\figno by1}
\else
\def\ifig#1#2#3{\xdef#1{fig.~\the\figno}
\writedef{#1\leftbracket fig.\noexpand~\the\figno}%
\global\advance\figno by1}
\fi

\font\cmss=cmss10 \font\cmsss=cmss10 at 7pt

\def\IB{\relax\hbox{$\inbar\kern-.3em{\rm B}$}}
\def\IC{\relax\hbox{$\inbar\kern-.3em{\rm C}$}}
\def\IQ{\relax\hbox{$\inbar\kern-.3em{\rm Q}$}}
\def\IO{\relax\hbox{$\inbar\kern-.3em{\rm O}$}}
\def\ID{\relax\hbox{$\inbar\kern-.3em{\rm D}$}}
\def\IE{\relax\hbox{$\inbar\kern-.3em{\rm E}$}}
\def\IF{\relax\hbox{$\inbar\kern-.3em{\rm F}$}}
\def\IG{\relax\hbox{$\inbar\kern-.3em{\rm G}$}}
\def\IGa{\relax\hbox{${\rm I}\kern-.18em\Gamma$}}
\def\IH{\relax{\rm I\kern-.18em H}}
\def\IK{\relax{\rm I\kern-.18em K}}
\def\IL{\relax{\rm I\kern-.18em L}}
\def\IP{\relax{\rm I\kern-.18em P}}
\def\IR{\relax{\rm I\kern-.18em R}}
\def\Z{\relax\ifmmode\mathchoice
{\hbox{\cmss Z\kern-.4em Z}}{\hbox{\cmss Z\kern-.4em Z}}
{\lower.9pt\hbox{\cmsss Z\kern-.4em Z}}
{\lower1.2pt\hbox{\cmsss Z\kern-.4em Z}}\else{\cmss Z\kern-.4em
Z}\fi}
\def\IZ{Z\!\!\!Z}
\def\II{\relax{\rm I\kern-.18em I}}

\def\S{{\bf S}}

\def\R{{\bf R}}

\def\CP{{\bf CP}}

\def\example#1{\bgroup\narrower\footnotefont\baselineskip\footskip\bigbreak
\hrule\medskip\nobreak\noindent {\bf Example}. {\it #1\/}\par\nobreak}
\def\endexample{\medskip\nobreak\hrule\bigbreak\egroup}

\def\CA {{\cal A}}

\def\CM {{\cal M}}
\def\CN {{\cal N}}
\def\CO {{\cal O}}
\def\CP {{\cal P}}

\def\CW {{\cal W}}


\def\p{\partial}

\def\tilde{\widetilde}
\def\hat{\widehat}
\def\bar{\overline}


\def\Tr{{\rm Tr}}

\def\Vol{{\rm Vol}}

\def\p{\partial}

\def\inbar{\,\vrule height1.5ex width.4pt depth0pt}
\def\r{{\rm Re}\ }
\def\i{{\rm Im}\ }

\def\a{\alpha}
\def\b{\beta}
\def\g{\gamma}

\def\om{\omega}

\def\bar{\overline}

\def\re{{\rm Re}\ }
\def\im{{\rm Im}\ }

\def\A{{\cal A}}
\def\N{{\cal N}}
\def\C{{\IC}}
\def\R{{\IR}}
\def\Z{{\IZ}}

\def\kahler{{K\"ahler}}
\def\ihalf{{i \over 2}}
\def\ifour{{i \over 4}}


\Title{\vbox{\baselineskip11pt\hbox{hep-th/0411073}
\hbox{HUTP-04/A042}
\hbox{ITFA-2004-54}
}} {\vbox{
\centerline{Topological M-theory as Unification}
\centerline{of Form Theories of Gravity}
}}
\centerline{
Robbert Dijkgraaf,$^1$
Sergei Gukov,$^2$
Andrew Neitzke,$^2$
and Cumrun Vafa$^2$}
\vskip 8pt
\centerline{\it $^1$ Institute for Theoretical Physics \&}
\centerline{\it Korteweg-de Vries Institute for Mathematics,}
\centerline{\it University of Amsterdam,}
\centerline{\it 1018 XE Amsterdam, The Netherlands}
\medskip
\centerline{\it $^2$
Jefferson Physical Laboratory, Harvard University,}
\centerline{\it Cambridge, MA 02138, USA}
\medskip
\medskip
\medskip
\noindent
We introduce a notion of topological M-theory and argue 
that it provides a unification of form theories of gravity in various dimensions.  
Its classical solutions involve $G_2$ holonomy metrics on 7-manifolds, obtained
from a topological action for a 3-form gauge field introduced by Hitchin. 
We show that by reductions of this 7-dimensional theory one can classically obtain 6-dimensional 
topological A and B models, the self-dual sector of loop quantum gravity in 4 dimensions,
and Chern-Simons gravity in 3 dimensions.  We also find
that the 7-dimensional M-theory perspective sheds some light on the fact that
the topological string partition function is a wavefunction,
as well as on S-duality between the A and B models.  The
degrees of freedom of the A and B models appear as conjugate variables
in the 7-dimensional theory.
Finally, from the topological M-theory perspective
we find hints of an intriguing holographic link between 
non-supersymmetric Yang-Mills in 4 dimensions and A model topological strings
on twistor space.

\smallskip
\Date{November 2004}

\listtoc\writetoc
\newsec{Introduction}

The search for a quantum theory of gravity has been a source of
puzzles and inspirations for theoretical physics over the past few
decades.  The most successful approach to date is string theory; but,
beautiful as it is, string theory has many extra aspects to it which
were not asked for.  These include the appearance of extra dimensions
and the existence of an infinite tower of increasingly massive
particles.  These unexpected features have been, at least in some
cases, a blessing in disguise; for example, the extra dimensions
turned out to be a natural place to hide the microstates of black
holes, and the infinite tower of particles was necessary in order for
the AdS/CFT duality to make sense.  Nevertheless, it is natural to ask
whether there could be simpler theories of quantum gravity.  If they
exist, it might be possible to understand them more deeply, leading us
to a better understanding of what it means to quantize gravity;
furthermore, simple theories of gravity might end up being the
backbone of the more complicated realistic theories of quantum
gravity.

In the past decade, some realizations of this notion of a ``simpler''
theory of gravity have begun to emerge from a number of different
directions.  The common thread in all these descriptions is that, in
the theories of gravity which appear, the metric is {\it not} one of
the fundamental variables.  Rather, these theories describe dynamics
of gauge fields or higher $p$-forms, in terms of which the metric can
be reconstructed.  These theories generally have only a finite number
of fields; we shall call them {\it form theories of gravity}.

Notable examples of form theories of gravity are\foot{One could also
include in this list, as will be discussed later in this paper, the
case of 2-dimensional gravity in the target space of the non-critical
$c=1$ string; in that case one gets a theory involving a symplectic
form on a 2-dimensional phase space, defining a Fermi surface, in term
of which the metric and other fields can be reconstructed.}  the
description of 3-dimensional gravity in terms of Chern-Simons theory,
the description of 4-dimensional gravity in terms of $SU(2)$ gauge
theory coupled to other fields, the description of the target space
theory of A model topological strings in terms of variations of the
\kahler\ 2-form, and the description of the target space theory of the
B model in terms of variations of the holomorphic 3-form.

Meanwhile, recent developments in the study of the topological A and B models
suggest that we need a deeper understanding of these theories.  On the one
hand, they have been conjectured to be S-dual to one another \refs{\NV,\NOV}.
On the other hand, the A model has been related to a quantum gravitational foam
\refs{\ORV, \INOV}.  Moreover, their nonperturbative definition has
begun to emerge through their deep connection with the counting of BPS
black hole states in 4 dimensions \refs{\OSV, \VafaYM}.  There is also
a somewhat older fact still in need of a satisfactory explanation: it
has been known for a while that the holomorphic anomaly of topological
strings \BCOV\ can be viewed as the statement that the partition
function of topological string is a state in some 7-dimensional
theory, with the Calabi-Yau 3-fold realized as the boundary of space
\Wittenhol\ (see also \DVV).

Parallel to the new discoveries about topological strings was the
discovery of new actions for which the field space consists of
``stable forms'' \Hitchin.  The critical points of these actions can
be used to construct special holonomy metrics.  A particularly
interesting example is a 3-form theory which constructs $G_2$ holonomy
metrics in 7 dimensions.  Interestingly enough, as we will explain,
the Hamiltonian quantization of this theory looks a lot like a
combination of the A and B model topological strings, which appear in
terms of conjugate variables.  All this hints at the existence of a
``topological M-theory'' in 7 dimensions, whose effective action leads
to $G_2$ holonomy metrics and which can reduce to the topological A
and B models.

The main aim of this paper is to take the first steps in developing a
unified picture of all these form theories of gravity.  Our aim is
rather modest; we limit ourselves to introducing some of the key ideas
and how we think they may be related, without any claim to presenting
a complete picture.  The 7-dimensional theory will be the unifying
principle; it generates the topological string theories as we just
noted, and furthermore, the interesting gravitational form theories in
3 and 4 dimensions can be viewed as reductions of this 7-dimensional
form theory near associative and coassociative cycles.

We will also find another common theme.  The form theories of gravity
naturally lead to calibrated geometries, which are the natural setting
for the definition of supersymmetric cycles where branes can be
wrapped.  This link suggests an alternative way to view these form
theories, which may indicate how to define them at the quantum level:
they can be understood as counting the BPS states of wrapped branes of
superstrings.  Namely, recall that in the superstring there is an
attractor mechanism relating the charges of the black hole (the
homology class of the cycle they wrap on) to specific moduli of the
internal theory (determining the metric of the internal manifold).  We
will see that the attractor mechanism can be viewed as a special case
of the general idea of obtaining metrics from forms.

The organization of this paper is as follows.  In Section 2, we
provide evidence for the existence of topological M-theory in 7
dimensions.  In particular, we use the embedding of the topological
string into the superstring to give a working definition of
topological M-theory in terms of topological strings in 6 dimensions,
with an extra circle bundle providing the ``11-th'' direction.  We
also give a more extensive discussion of how the very existence of
topological M-theory could help resolve a number of puzzles for
topological strings in 6 dimensions.  In Section 3, we give a short
review of some form gravity theories in dimensions 2, 3, 4 and 6.  In
Section 4, we discuss some new action principles constructed by
Hitchin, which lead to effective theories of gravity in 6 and 7
dimensions.  These gravity theories are related to special holonomy
manifolds and depend on the mathematical notion of ``stable form,'' so
we begin by reviewing these topics; then we introduce Hitchin's
actions in 6 and 7 dimensions, as well as a classical Hamiltonian
formulation of the 7-dimensional theory.  In Section 5, we argue that
these new gravity theories in 6 dimensions are in fact reformulations
of the target space dynamics of the A and B model topological string
theories.  In Section 6, we show how the 7-dimensional theory reduces
classically to the 3, 4 and 6-dimensional gravity theories we reviewed
in Section 3.  In Section 7, we discuss canonical quantization of the
7-dimensional theory; we show that it is related to the A and B model
topological strings, and we argue that this perspective could shed
light on the topological S-duality conjecture.  In Section 8, we
reinterpret the gravitational form theories as computing the entropy
of BPS black holes.  In Section 9, we discuss a curious holographic
connection between twistor theory and the topological $G_2$ gravity.
In Section 10 we discuss possible directions for further development
of the ideas discussed in this paper.  Finally, in an appendix, we
discuss an interesting connection between the phase space of
topological M-theory and ${\cal N}=1$ supersymmetric vacua in 4
dimensions.

\newsec{Evidence for Topological M-theory}

In order to define a notion of topological M-theory, we exploit the
connection between the physical superstring and the physical M-theory.
Recall that we know that topological strings make sense on Calabi-Yau
3-folds, and topological string computations can be embedded into the
superstring.  It is natural to expect that the dualities of the
superstring, which found a natural geometric explanation in M-theory,
descend to some dualities in topological theories, which might find a
similar geometric explanation in topological M-theory.  Thus a natural
definition of topological M-theory is that it should be a theory with
one extra dimension relative to the topological string, for a total of
7.  Moreover, we should expect that {\it M-theory on} $M \times \S^1$
{\it is equivalent to topological strings on} $M$, where $M$ is a
Calabi-Yau manifold.  More precisely, here we are referring to the
topological A model on $M$.  The worldsheets of A model strings are
identified with M-theory membranes which wrap the $\S^1$.  Later we
will see that in some sense the M-theory formalism seems to
automatically include the B model along with the A model, with the two
topological string theories appearing as conjugate variables.  The
topological string should be a good guide to the meaning of
topological M-theory, at least in the limit where the $\S^1$ has small
radius.  One would expect that the radius of the $\S^1$ gets mapped to
the coupling constant of the topological string. Of course,
topological M-theory should provide an, as yet not well-defined,
nonperturbative definition of topological string theory.

So far we only discussed a constant size $\S^1$, but we could also
consider the situation where the radius is varying, giving a more
general 7-manifold.  The only natural class of such manifolds which
preserves supersymmetry and is purely geometric is the class of $G_2$
holonomy spaces; indeed, that there should be a topological theory on
$G_2$ manifolds was already noted in \HM, which studied Euclidean
M2-brane instantons wrapping associative 3-cycles.  So consider
M-theory on a $G_2$ holonomy manifold $X$ with a $U(1)$ action.  This
is equivalent to the Type IIA superstring, with D6 branes wrapping
Lagrangian loci on the base where the circle fibration degenerates.
{\it We define topological M-theory on $X$ to be equivalent to A model
topological strings on $X / U(1)$, with Lagrangian D-branes inserted
where the circle fibration degenerates}.  This way of defining a
topological M-theory on $G_2$ was suggested in \refs{\AKV,\AV}.

In this setting, the worldsheets of the A model can end on the
Lagrangian branes; when lifted up to the full geometry of $X$ these
configurations correspond to honest closed 3-cycles which we identify
as membrane worldvolumes.  Moreover, string worldsheets which happen
to wrap {\it holomorphic} cycles of the Calabi-Yau lift to membranes
wrapping associative 3-cycles of the $G_2$ holonomy manifold.  So,
roughly speaking, we expect that topological M-theory should be
classically a theory of $G_2$ holonomy metrics, which gets quantum
corrected by membranes wrapping associative 3-cycles --- in the same
sense as the topological A model is classically a theory of \kahler\
metrics, which gets quantum corrected by strings wrapping holomorphic
cycles.  We can be a little more precise about the coupling between
the membrane and the metric: recall that a $G_2$ manifold comes
equipped with a 3-form $\Phi$ and a dual 4-form $G=*\Phi$, in terms of
which the metric can be reconstructed.  We will see that it is natural
to consider this $G$ as a field strength for a gauge potential,
writing $G=G_0+d\Gamma$; then $\Gamma$ is a 3-form under which the
membrane is charged.

So we have a workable definition of topological M-theory, which makes
sense on 7-manifolds with $G_2$ holonomy, at least perturbatively to
all orders in the radius of the circle.  Thus the existence of the
theory is established in the special cases where we have a $U(1)$
action on $X$; we conjecture that this can be extended to a theory
which makes sense for arbitrary $G_2$ holonomy manifolds.  This is
analogous to what we do in the physical superstring; we do not have an
{\it a priori} definition of M-theory on general backgrounds, but only
in special situations.

Now that we have established the existence of a topological M-theory
in 7 dimensions (more or less at the same level of rigor as for the
usual superstring/M-theory relation), we can turn to the question of
what new predictions this theory makes.  Indeed, we now suggest that
it may solve two puzzles which were previously encountered in the
topological string.

There has been a longstanding prediction of the existence of a
7-dimensional topological theory from a very different perspective,
namely the wavefunction property of the topological string partition
function, which we now briefly recall in the context of the B model.
The B model is a theory of variations $\delta \Omega$ of a holomorphic
3-form on a Calabi-Yau 3-fold $X$.  Its partition function is written
$Z_B(x;\Omega_0)$.  Here $x$ refers to the zero mode of $\delta
\Omega$, $x \in H^{3,0}(X,\C) \oplus H^{2,1}(X,\C)$, which is not
integrated over in the B model.  The other variable $\Omega_0$ labels
a point on the moduli space of complex structures on $X$; it specifies
the background complex structure about which one perturbs.  Studying
the dependence of $Z_B$ on $\Omega_0$ one finds a ``holomorphic
anomaly equation'' \refs{\BCOV,\BCOVa}, which is equivalent to the
statement that $Z_B$ is a {\it wavefunction} \Wittenhol, defined on
the phase space $H^3(X,\R)$.  Namely, different $\Omega_0$ just
correspond to different polarizations of this phase space, so
$Z_B(x;\Omega_0)$ is related to $Z_B(x;\Omega'_0)$ by a Fourier-type
transform.  This wavefunction behavior is mysterious from the point of
view of the 6-dimensional theory on $X$.  On the other hand, it would
be natural in a 7-dimensional theory: namely, if $X$ is realized as
the boundary of a 7-manifold $Y$, then path integration over $Y$ gives
a wavefunction of the boundary conditions one fixes on $X$.

Another reason to expect a 7-dimensional description comes from the
recent conjectures that the A model and B model are independent only
perturbatively.  Namely, each contains nonperturbative objects which
could naturally couple to the fields of the other.  The branes in the
A model are wrapped on Lagrangian cycles, the volume of which are
measured by some 3-form, and it is natural to identify this 3-form
with the holomorphic 3-form $\Omega$ of the B model; conversely, the
branes in the B model are wrapped on holomorphic cycles, whose volumes
would be naturally measured by the \kahler\ form $k$ of the A model.
This observation has led to the conjecture \refs{\NV,\NOV} that
nonperturbatively both models should include both types of fields and
branes, and in fact that the two could even be S-dual to one another.
One is thus naturally led to search for a nonperturbative formulation
of the topological string which would naturally unify the A and B
model branes and fields.  Such a unification is natural in the
7-dimensional context: near a boundary with unit normal direction
$dt$, the 3- and 4-forms $\Phi,G$ defining the $G_2$ structure
naturally combine the fields of the A and B model on the boundary,
\eqn\phiano{\eqalign{
& \Phi = \re \Omega + k \wedge dt, \cr
& G = \im \Omega \wedge dt + \half k\wedge k.}}

Later we will see that the A and B model fields are canonically
conjugate in the Hamiltonian reduction of topological M-theory on
$X\times \R$.  In particular, the wavefunctions of the A model and the
B model cannot be defined simultaneously.

\newsec{Form Theories of Gravity in Diverse Dimensions}

The long-wavelength action of the ``topological M-theory'' we are
proposing will describe metrics equipped with a $G_2$ structure.  In
fact, as we will discuss in detail, the 7-dimensional metric in this
theory is reconstructed from the 3-form $\Phi$ (or equivalently, from
the 4-form $G=*\Phi$).  This might at first seem exotic: the metric is
not a fundamental field of this theory but rather can be reconstructed
from $\Phi$.  However, similar constructions have appeared in lower
dimensions, where it is believed at least in some cases that the
reformulation in terms of forms (``form theory of gravity'') is a
better starting point for quantization: we know how to deal with gauge
theories, and perhaps more general form theories, better than we know
how to deal with gravity.  Of course, {\it rewriting a classical
theory of gravity in terms of classical forms is no guarantee that the
corresponding quantum theory exists}.  We are certainly not claiming
that arbitrary form theories make sense at the quantum level!

Nevertheless, in low dimensions some special form gravity theories
have been discussed in the literature, which we believe do exist in
the quantum world --- and moreover, as we will see, these theories are
connected to topological M-theory, which we have already argued should
exist.

In this section we review the form gravity theories in question.  They
describe various geometries in $2$, $3$, $4$ and $6$ dimensions.  Here
we will discuss mainly their classical description.  It is more or
less established that the theories discussed below in dimensions $2$,
$3$, and $6$ exist as quantum theories, at least perturbatively.  In
dimensions $2$ and $6$, this is guaranteed by topological string
constructions.  In dimension $3$ also, the quantum theory should exist
since it is known to lead to well defined invariants of 3-manifolds.
The 4-dimensional theory, which gives self-dual gravity, is not known
to exist in full generality, although for zero cosmological constant,
it is related to the Euclidean $\N=2$ string, which is known to exist
perturbatively \OV.  For the case of nonzero cosmological constant, we
will give further evidence that the theory exists at the quantum level
by relating it to topological M-theory later in this paper.

\subsec{$2D$ Form Gravity}

By the $2D$ form theory of gravity we have in mind the theory which
appears in the target space of non-critical bosonic strings, or more
precisely, the description of the large $N$ limit of matrix models in
terms of the geometry of the eigenvalue distribution.  The basic idea
in this theory of gravity is to study fluctuations of a Fermi surface
in a 2-dimensional phase space.  The dynamical object is the area
element $\omega$, representing the phase space density, which is
defined to be non-zero in some region $R$ and vanishing outside.  By a
choice of coordinates, we can always write this area element as
$\omega=dx\wedge dp$ inside $R$ and zero elsewhere.  Hence the data of
the theory is specified by the boundary $\partial R$, which we can
consider locally as the graph of a function $p(x)$.  The study of the
fluctuations of the boundary is equivalent to that of fluctuations of
$\omega$.  Actually, in this gravity theory one does not allow
arbitrary fluctuations; rather, one considers only those which
preserve the integral
\eqn\preserveint{\oint p(x)\,dx=A.}
Such fluctuations can be written $p(x)=p_0(x) +\partial \phi(x)$.  In other words,
the cohomology class of $\omega$, or ``zero mode,'' is fixed by $A$, and the ``massive modes'' 
captured by the field $\phi(x)$ are the dynamical degrees of freedom.

This gravity theory is related to the large $N$ limit of matrix
models, where $x$ denotes the eigenvalue of the matrix and $p(x)\,dx$
denotes the eigenvalue density distribution.  $A$ gets interpreted as
the rank $N$ of the matrix.  One can solve this theory using matrix
models, or equivalently using $W_\infty$ symmetries \ADKMV.

This theory can also be viewed \refs{\ADKMV,\AAMV} as the effective
theory of the B model topological string on the Calabi-Yau 3-fold
$uv=F(x,p)$, where $F(x,p)=0$ denotes the Fermi surface.  In this
language, $\omega = dx \wedge dp$ is the reduction of the holomorphic
3-form to the $(x,p)$ plane.  In the B model one always fixes the
cohomology class of the 3-form; here this reduces to fixing the area
$A$ as we described above.

\subsec{$3D$ Gravity Theory as Chern-Simons Gauge Theory}

Now we turn to the case of three dimensions.
Pure gravity in three dimensions is topological,
in a sense that it does not have propagating gravitons.
In fact, we can write the usual Einstein-Hilbert action
with cosmological constant $\Lambda$,
\eqn\sgravgr{ S_{{\rm grav}} = \int_M \sqrt{g} \Big( R - 2 \Lambda \Big), }
in the first-order formalism,
\eqn\sgravee{ S = \int_M
\Tr \Big( e \wedge F + {\Lambda \over 3} e \wedge e \wedge e \Big), }
where $F=d A + A \wedge A$ is the field strength of an $SU(2)$ gauge
connection $A^i$, and $e^i$ is an $SU(2)$-valued 1-form on $M$.
Notice that the gravity action \sgravee\ has the form of a BF theory,
and does not involve a metric on the 3-manifold $M$.  A metric (of
Euclidean signature) can however be reconstructed from the fundamental
fields --- namely, given the $SU(2)$-valued 1-form $e$, one can write
\eqn\gtriviae{ g_{ab} = - \half \Tr (e_a e_b). }

The equations of motion that follow from \sgravee\
are:
\eqn\eeeeom{\eqalign{
& D_A e = 0, \cr
& F + \Lambda e \wedge e = 0.
}}
The first equation says that $A$ can be interpreted
as the spin connection for the vielbein $e^i$,
while the second equation is the Einstein equation
with cosmological constant $\Lambda$.

Gravity in three dimensions has a well-known reformulation
in terms of Chern-Simons gauge theory \refs{\AT,\Wittengrav}, 
\eqn\csaction{ S = \int_M \Tr \Big( \A \wedge d\A + {2 \over 3} \A
\wedge \A \wedge \A \Big), }
where $\A$ is a gauge connection with values in
the Lie algebra of the gauge group $G$.
The gauge group $G$ is determined by the cosmological
constant, and can be viewed as the isometry group
of the underlying geometric structure.
Specifically, in the Euclidean theory, it is either
$SL(2,\IC)$, or $ISO(3)$, or $SU(2) \times SU(2)$,
depending on the cosmological constant:

\vskip 0.8cm
\vbox{
\centerline{\vbox{
\hbox{\vbox{\offinterlineskip
\def\tablespace{height7pt&\omit&&\omit&&\omit&&\omit&\cr}
\def\tablerule{\tablespace\noalign{\hrule}\tablespace}

\hrule\halign{&\vrule#&\strut\hskip0.2cm\hfill
#\hfill\hskip0.2cm\cr
\tablespace & Cosmological Constant && $\Lambda < 0$ && $\Lambda=0$ && $\Lambda > 0$ &\cr
\tablerule & Gauge group $G$ && $SL(2,\IC)$ && $ISO(3)$ && $SU(2) \times SU(2)$ &\cr
\tablespace}\hrule}}}}
\centerline{ \hbox{{\bf Table 1:}{\it ~~ Euclidean $3D$ gravity can
be viewed as a Chern-Simons theory.}}}
} \vskip 0.5cm

The equations of motion that follow from the
Chern-Simons action \csaction\ imply that the gauge
connection $\A$ is flat,
\eqn\cseom{ d\A + \A \wedge \A = 0 }
Writing this equation in components one can reproduce
the equations of motion \eeeeom.
For example, if $\Lambda < 0$ one can write the complex
gauge field $\A$ as $\A^k = w^k + i e^k$. Substituting this
into \cseom\ and combining the real and imaginary terms,
we recognize the equations \eeeeom\ with $\Lambda = -1$.
Finally, we note that, in the Chern-Simons theory, the gauge
transformation with a parameter $\epsilon$ has the form
\eqn\csgauge{ \delta_{\epsilon} \A = d \epsilon - [\A, \epsilon]. }

One can also describe the quantum version of
$3D$ gravity directly via various discrete models.
For example, given a triangulation $\Delta$ of $M$
one can associate to each tetrahedron a quantum $6j$-symbol
and, following Turaev and Viro \TV, take the state sum
\eqn\tvdef{ TV (\Delta) =
\Big( - {(q^{1/2} - q^{-1/2})^2 \over 2k} \Big)^V
\sum_{j_e} \prod_{{\rm edges}} [2j_e + 1]_q
\prod_{{\rm tetrahedra}} (6j)_q }
where $V$ is the total number of vertices in the triangulation,
and $[2j + 1]_q$
is the quantum dimension of the spin $j$ representation of $SU(2)_q$
\eqn\qdimdef{
[n]_q = {q^{n/2} - q^{-n/2} \over q^{1/2} - q^{-1/2} } }
One can prove \TV\ that the Turaev-Viro invariant is independent on
the triangulation and, therefore, gives a topological invariant,
$TV(M) = TV(\Delta)$.  Furthermore, it has been shown by Turaev
\Turaev\ and Walker \Walker\ that the Turaev-Viro invariant is equal
to the square of the partition function in $SU(2)$ Chern-Simons theory
(also known as the Reshetikhin-Turaev-Witten invariant
\refs{\WittenJones,\RT}):
\eqn\tvzsutwo{TV(M) = \vert Z_{SU(2)}(M) \vert^2}
There is a similar relation between the
$SL(2,\IC)$ Chern-Simons partition function
and quantum invariants of hyperbolic 3-manifolds \Apol.

\subsec{$4D$ 2-Form Gravity}

In dimension four, there are several versions of
``topological gravity''.
Here we review a theory known as 2-form gravity
\refs{\Plebanski,\Horowitz,\torrea,\torreb,\ANU,\LNU,\NakamichiI,\Abe}, which also describes
the self-dual sector of loop quantum gravity \Smolin.

We begin by writing the action for Einstein's
theory in a slightly unconventional way \refs{\Israel,\Plebanski,\CDJM}:
\eqn\tfaction{ 
S_{4D} = \int_{M^4} \Sigma^k \wedge F_k
- {\Lambda \over 24} \Sigma^k \wedge \Sigma_k + \Psi_{ij} \Sigma^i
\wedge \Sigma^j.
}
Here $A^k$ is an $SU(2)$ gauge field, with curvature 
$F^k = dA^k + \epsilon^{ijk} A^j \wedge A^k$, 
and $\Sigma^k$ is an $SU(2)$ triplet of 2-form fields, $k=1,2,3$.
The parameter $\Lambda$ will be interpreted below as a cosmological
constant.  Finally, $\Psi_{ij} = \Psi_{(ij)}$ is a scalar field on
$M$, transforming as a symmetric tensor of $SU(2)$.

To see the connection to ordinary general relativity, 
one constructs a metric out of the two-form field $\Sigma^k$ as follows.
The equation of motion from varying 
$\Psi_{ij}$ implies that $\Sigma^k$ obeys the constraint
\eqn\ssconstr{ \Sigma^{(i} \wedge \Sigma^{j)}
- {1 \over 3} \delta^{ij} \Sigma_k \wedge \Sigma^k = 0.}
When \ssconstr\ is satisfied 
the two-form $\Sigma^k$ may be reexpressed in terms of
a vierbein \CDJM,
\eqn\sviaee{ \Sigma^k = - \eta^k_{ab} e^a \wedge e^b.}
Here $e^a$ are vierbein 1-forms on $M^4$, $a=1,\ldots,4$,
and $\eta^k_{ab}$ is the `t Hooft symbol,
\eqn\thooftsymb{
\eta^k_{ab} = \epsilon^k_{ab0} + \half \epsilon^{ijk} \epsilon_{ijab}.
}
In other words,
$$\Sigma^1=e^{12}-e^{34},$$
$$\Sigma^2=e^{13}-e^{42},$$
$$\Sigma^3=e^{14}-e^{23},$$
where $e^{ij}=e^i\wedge e^j$.
The vierbein in turn determines a metric in the usual way:
\eqn\metricG{ g = \sum_{a=1}^4 e^a \otimes e^a. }
With respect to this metric $g$, the two-forms $\Sigma^k$ ($k = 1, 2,
3$) are all self-dual in the sense that $\Sigma^k = * \Sigma^k$.
(This just follows from their explicit expression \sviaee\ in 
terms of the vierbein.)

One can also write the metric directly in terms of $\Sigma$
without first constructing the vierbein \refs{\CDJM,\Urbantke}:
\eqn\gviasss{
\sqrt{g}\,g_{ab} = - {1 \over 12} \Sigma^i_{a a_1} \Sigma^j_{b a_2}
\Sigma^k_{a_3 a_4} \epsilon^{ijk} \epsilon^{a_1 a_2 a_3 a_4}.
}

Having constructed the metric $g$ out of $\Sigma$, we now want to check that it obeys
Einstein's equation on-shell.  The equations of motion which follow from
\tfaction\ are
\eqn\eomfour{\eqalign{
& D_A \Sigma  = 0, \cr
& F_i  = \Psi_{ij} \Sigma^j + {\Lambda \over 12} \Sigma_i.
}}
The first equation $D_A \Sigma = 0$ says that $A$ is the self-dual
part of the spin connection defined by the metric $g$.
The second equation then contains information about the Riemann
curvature\foot{Here we are
considering $R$ with all indices down, $R_{abcd}$, as a symmetric map
$\Lambda^2 \to \Lambda^2$.} acting on self-dual 2-forms $\Lambda^2_+$.
Namely, since the $\Sigma^k$ appearing on the right side 
are also self-dual two-forms, the 
Riemann curvature maps $\Lambda^2_+ \to \Lambda^2_+$.
By the standard decomposition of the Riemann tensor, this implies
that the trace-free part of the Ricci curvature vanishes.
Then $\Psi_{ij}$ is identified with the self-dual part
of the Weyl curvature, and the last term gives 
the trace part of the Ricci tensor,
consistent with the cosmological constant $\Lambda$.

So far we have seen that the action \tfaction\ reproduces
Einstein's theory of gravity, in the sense that the classical
solutions correspond exactly to Einstein metrics on $M$ with
cosmological constant $\Lambda$.  Now let us consider the 
effect of dropping the field $\Psi_{ij}$, giving
\eqn\tfactionsd{ 
S_{4D} = \int_{M^4} \Sigma^k \wedge F_k
- {\Lambda \over 24} \Sigma^k \wedge \Sigma_k.
}
One can consider \tfactionsd\ as obtained by starting from
\tfaction, multiplying the last term by $\epsilon$, and then
taking the $\epsilon \to 0$ limit.  Just when we reach $\epsilon = 0$ we seem 
to lose the constraint \ssconstr, which was the 
equation of motion for $\Psi_{ij}$ and was crucial for the description
of $\Sigma^k$ in terms of the vierbein.  However, at $\epsilon = 0$
something else happens:  the action develops a large new symmetry,
\eqn\newsymm{\eqalign{
\delta A_k &= {\Lambda \over 12}{\theta_k}, \cr
\delta \Sigma_k &= D_A \theta_k.
}}
This new symmetry can be used to reimpose the constraint \ssconstr,
so in this sense the $\epsilon \to 0$ limit is smooth and sensible to
consider.  The only change to the equations of motion
is that the term $\Psi_{ij} \Sigma^j$ disappears from the right side 
of \eomfour, leaving
\eqn\eomfoursd{\eqalign{
& D_A \Sigma  = 0, \cr
& F_i = {\Lambda \over 12} \Sigma_i.
}}
As we mentioned above, the $\Psi_{ij}$ term represents the
self-dual part of the Weyl curvature; so \eomfoursd\ imply that 
the metric constructed from $\Sigma$ is not only Einstein but also has
vanishing self-dual Weyl curvature.
In this sense the action \tfactionsd\
gives rise to anti-self-dual Einstein manifolds,
\eqn\rweom{ R_{ab} = \Lambda g_{ab}, \quad\quad W^{(+)}_{abcd}=0. }
Note that such manifolds are rather rare compared
to ordinary Einstein manifolds;
for example, with $\Lambda > 0$ there are just
two smooth examples, namely ${\bf CP}^2$ and ${\bf S}^4$ \Hittwistor.
With $\Lambda = 0$ the solutions are hyperk\"ahler metrics
in 4 dimensions; these are target spaces for the $\N=2$ string (or
equivalently the $\N=4$ topological string), which  
provides a completion of the self-dual gravity
theory in that case.

\subsec{$6D$ Form Theories: \kahler\ and Kodaira-Spencer Gravity}

In dimension 6, two different form theories of gravity
arise in ($\N=2$) topological string theory.
One, known as the \kahler\ gravity theory \BSadov,
describes the target space gravity (string field theory)
of the topological A model.
The second theory, called the Kodaira-Spencer theory of gravity \BCOV,
is the string field theory of the topological B model
and describes variations of the complex structure.
Below we review each of these theories in turn.

We begin with the B model.
The basic field of the Kodaira-Spencer gravity theory is a vector-valued
1-form field $A$, for which the action is given by \BCOV
\eqn\ksaction{
S_{{\rm KS}} = {1 \over 2} \int_M A' {1 \over \p} \bar \p A'
+ {1 \over 6} \int_M (A \wedge A)' \wedge A'.
}
Here, we use the standard notation $A' := (A \cdot \Omega_0)$
for the product with the background holomorphic $(3,0)$ form.  The field
$A$ then defines a variation of $\Omega$, given by the
formula
\eqn\omviaa{\Omega = \Omega_0 + A' + (A \wedge A)' + (A \wedge A \wedge A)'.}
This expression for the variation of $\Omega$
follows from its local expression in complex coordinates,
$\Omega = \Omega_{ijk} dz^i \wedge dz^j \wedge dz^k$,
where $A$ is interpreted as giving a variation of the 
1-form $dz^i$:
\eqn\dzchng{ dz^i \mapsto dz^i + A_{\bar j}^i d \bar z^{\bar j}. }
In order that the ``non-local term'' ${1 \over \p} \bar{\p} A'$ in the
action \ksaction\ make sense, $A$ is not allowed to be an arbitrary
vector-valued 1-form; rather, there is a constraint
\eqn\ksconstr{
\p A' = 0.
}
Using this constraint we write 
\eqn\aflddecomp{ A' = x + \p \phi,
\quad \quad \quad x \in H^{2,1} (M,\C). }
Here the harmonic $x \in H^{2,1} (M,\C)$ represents the massless modes
(moduli) of $\Omega$, which are frozen in the Kodaira-Spencer theory,
while $\phi \in \Omega^{1,1}(M,\C)$ represents the massive modes,
which are the dynamical degrees of freedom.  Substituting \aflddecomp\
into \ksaction, we can write the Kodaira-Spencer action without
non-local terms:
\eqn\ksactionphi{
S_{{\rm KS}} = {1 \over 2} \int_M \p \phi \wedge \bar \p \phi
+ {1 \over 6} \int_M (A \wedge A)' \wedge A'
}
The equation of motion that follows from the action \ksactionphi\
is
\eqn\kseom{ \bar \p A' + \p (A \wedge A)' =0. }
Using \ksconstr\ and \kseom\ together one finds that the holomorphic
3-form \omviaa\ is closed on-shell,
\eqn\ksomeom{ d \Omega = 0. }
Hence solutions to the Kodaira-Spencer field equations correspond
to deformations $\Omega$ of the holomorphic 3-form $\Omega_0$.

When we view $\phi$ as the dynamical degree of freedom we must note
that it has a large shift symmetry, 
\eqn\shiftsymm{\phi \mapsto \phi+ \epsilon,}
where $\partial \epsilon=0$.  This symmetry can be used to set the
anti-holomorphic part of $\phi$ to zero, i.e.  $\overline \partial
\phi=0$.  In other words, $\phi$ should be viewed as the analog of the
{\it chiral} boson in 2 dimensions; in this sense the Kodaira-Spencer
theory is really a chiral theory.  In fact, in the local geometry we
discussed in Section 3.1, $\phi$ gets identified with a chiral boson
on the Riemann surface $F(x,p)=0$.

Although $A'$ and the Kodaira-Spencer action depend on $\partial \phi$
rather than on $\phi$ itself, it turns out that D1-branes of the B
model are charged under $\phi$ directly.  To see this, consider a
D1-brane wrapped on a 2-cycle $E$ which moves to another 2-cycle $E'$.
There is a 3-chain $C$ which interpolates between $E$ and $E'$, and
the variation of the action is given by (absorbing the string coupling
constant into $\Omega$)
\eqn\varaction{
\delta S=\int_{C} \Omega =\int_{C}\partial \phi= \int_{C} d\phi =
\int_{E}\phi - \int_{E'}\phi.
}
This coupling also explains the fact that a D1-brane is a source for
$\Omega$ (and hence shifts the integral of $\Omega$ on a 3-cycle
linking it.)  Namely, including such a source localized along $E$
would modify the equations of motion to \ADKMV
\eqn\opaa{
\overline \partial A'={\overline \partial} \partial \phi =\delta^4_E,
}
so that the kinetic term $\phi \partial {\overline \partial} \phi$ 
from \ksactionphi\ becomes precisely $\int_E\phi$.

The fact that D1-branes couple to $\phi$ has an important consequence:
there is an extra $H^{1,1}(M,\C)$ worth of degrees of freedom in
$\phi$, corresponding to the freedom to shift $\phi$ by a harmonic
form $b$, which does not affect $A'$ but does figure into
nonperturbative aspects of the B-model.  Namely, the amplitudes
involving D1-brane instantons, which should presumably be included in
the nonperturbative definition of the B model, are sensitive to this
shift.  Thus the partition function of the B model is
nonperturbatively a function both of $x \in H^{2,1}(M)$ and of $b \in
H^{1,1}(M,\C)$.  The necessity of the extra field $b$ was also
recently noted in \WB. 

As we will discuss later in more detail, it is natural that in a
nonperturbative definition the periods of $\Omega$ are quantized in
units of $g_s$. There is an overall $1/g_s^2$ in front of the closed
string action, so this will then give the appropriate $1/g_s$ coupling
of the field $\phi$ to the D-branes. Because of this flux
quantization, the corresponding ``Wilson lines'' $b$ will be naturally
periodic or $\C^*$ variables.

Having discussed the Kodaira-Spencer theory, let us now turn to
another 6-dimensional form theory of gravity, namely the K\"ahler
gravity theory, which describes variations of the K\"ahler structure
on $M$.  Its action is \refs{\BSadov,\Wittencs}
\eqn\kgravact{ S_{{\rm Kahler}} = \int_M
\Big( {1 \over 2} K {1 \over d^{c \dagger}} d K
+ {1 \over 3} K \wedge K \wedge K \Big), }
where $K$ is a variation of the (complexified) K\"ahler form
on $M$, and $d^c = \p - \bar{\p}$.
The K\"ahler gravity action \kgravact\ is
invariant under gauge transformations of the form
\eqn\kgravgauge{ \delta_{\a} K = d \a - d^{c \dagger} (K \wedge
\a), }
where $\a$ is a 1-form on $M$, such that $d^{c \dagger} \a =0$.
The equations of motion in the K\"ahler gravity theory are
\eqn\kgraveom{ d K + d^{c \dagger} (K \wedge K) = 0. }
As in the Kodaira-Spencer theory, we can decompose $K$
into massless and massive modes,
\eqn\kddecomp{ K = x + d^{c \dagger} \g,
\quad \quad \quad x \in H^{1,1} (M,\C) }
where $x \in H^{1,1}(M,\C)$ represents the K\"ahler moduli, which are
not integrated over, and $\g \in \Omega^3(M)$ contains the massive
modes of $K$.  Using \kddecomp, we can write the K\"ahler gravity
action \kgravact\ without non-local terms,
\eqn\kgravactg{ S_{{\rm Kahler}} = \int_M
\Big( {1\over 2} d \g \wedge d^{c \dagger} \g
+ {1 \over 3} K \wedge K \wedge K \Big). }

Just as in the B model, Lagrangian D-branes of the A model are charged
under $\g$, implying that these branes are sources for $K$ and hence
modify the integral of $K$ on 2-cycles which link them.  This also
implies that the partition function of the A model depends
nonperturbatively on the choice of a cohomology class in $H^3(M)$ as
well as on $x\in H^2(M)$.  Here the same remarks about flux quantization
hold as in the B model.

\newsec{Hitchin's Action Functionals}

In the previous section, we discussed various form theories of gravity 
which have appeared previously in the physics literature.  
Now we turn to some new ones.  We will describe actions constructed by
Hitchin \refs{\Hitchin,\Hit} for which the equations of motion yield
special geometric structures: either holomorphic 3-forms $\Omega$ and
symplectic structures $k$ in 6 dimensions, or $G_2$ holonomy metrics
in 7 dimensions.  As for the form theories we considered above, the
classical fields in these theories are real $p$-forms, from which the
desired geometric structures can be reconstructed.  In 6 dimensions
one has a 3-form $\rho$ and a 4-form $\sigma$; these forms will be
interpreted as $\rho = \re \Omega$, $\sigma = \half k \wedge k$. In 7
dimensions, one just has a single 3-form $\Phi$ (or its dual 4-form
$G$), interpreted as the associative 3-form (resp. coassociative
4-form) of the $G_2$ metric.

These action functionals have been used in the physics
literature to construct new metrics with special holonomy
\refs{\GYZ,\CCGLPW,\CGLP}. In the present context, they should
be regarded as effective actions for gravity theories.
In 6 dimensions, we will see in Section 5 that these gravity
theories are related to topological strings.  The 7-dimensional
action defines a new gravity theory which we identify as 
the low energy limit of topological M-theory.

\ifig\critptsfig{Critical points of volume functionals define
special geometric structures on $X$.}
{\epsfxsize3.5in\epsfbox{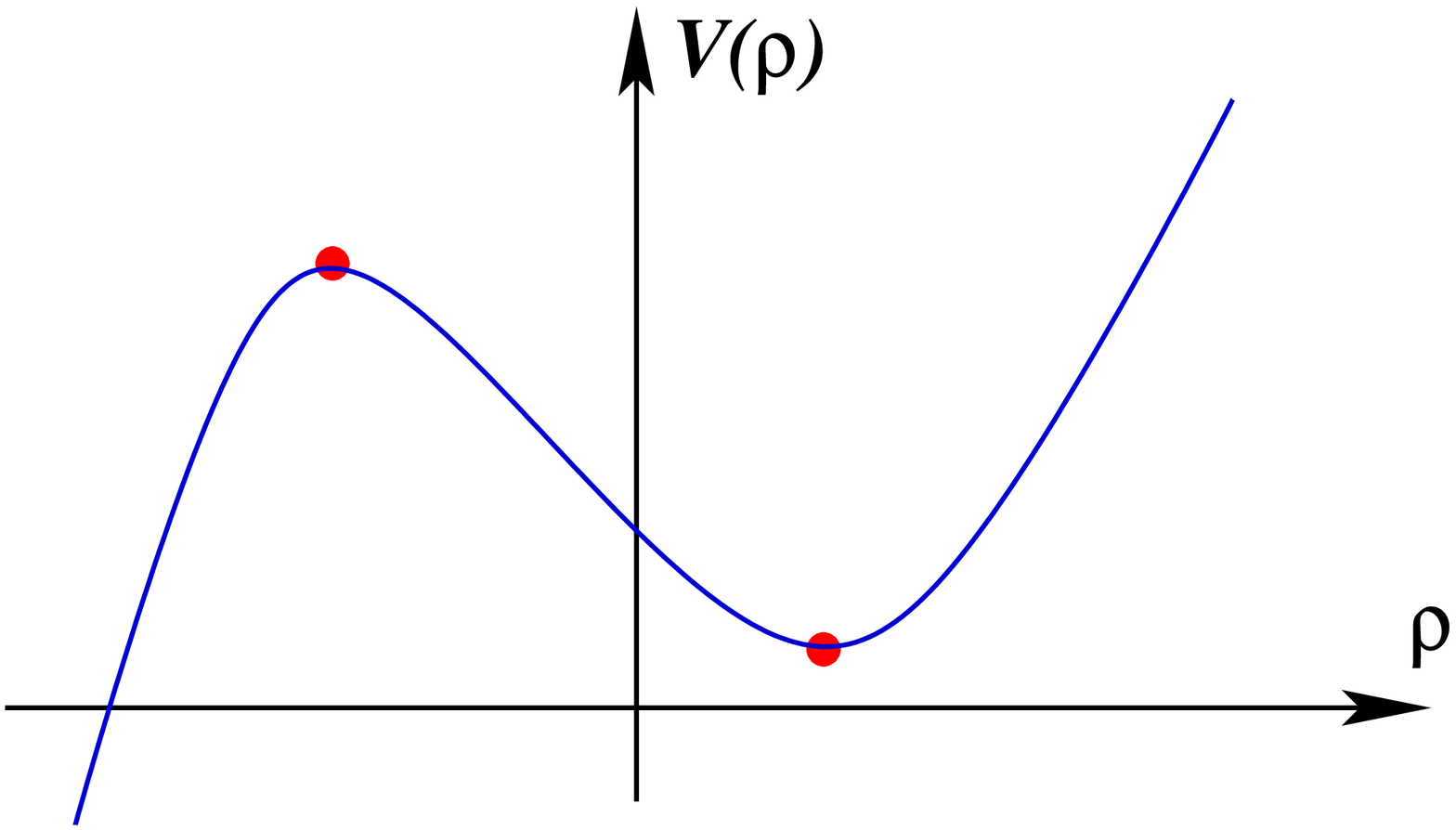}}

\subsec{Special Holonomy Manifolds and Calibrations}

In this subsection we briefly review the notion of special holonomy,
which plays an important role in supersymmetric string
compactifications, and which we expect to be important for topological
string/membrane theories. In particular, we emphasize that the
geometric structure on a special holonomy manifold $X$ can be
conveniently characterized by certain $p$-forms, invariant under the
holonomy group, ${\rm Hol} (X)$.

Recall that for any $n$-dimensional Riemannian manifold
$X$ we have 
\eqn\holson{ {\rm Hol} (X) \subseteq SO(n). }
The manifolds with special (reduced) holonomy are characterized
by the condition ${\rm Hol} (X) \subset SO(n)$.
Below we list some examples of special holonomy manifolds
that will be relevant in what follows.

\vskip 0.8cm
\vbox{
\centerline{\vbox{
\hbox{\vbox{\offinterlineskip
\def\tablespace{height7pt&\omit&&\omit&&\omit&&\omit&&\omit&\cr}
\def\tablerule{\tablespace\noalign{\hrule}\tablespace}

\hrule\halign{&\vrule#&\strut\hskip0.2cm\hfill
#\hfill\hskip0.2cm\cr
\tablespace
&  Metric && Holonomy && $n$ && SUSY && Invariant $p$-forms &\cr
\tablerule
&  Calabi-Yau && $SU(3)$ && $6$ && $1/4$ && $p=2$: ~$K$~~~(K\"ahler) &\cr
&   &&  &&  &&  && $p=3$: ~$\Omega$~~~~~~~~~~~~~~~ &\cr
\tablerule
&  Exceptional && $G_2$ && $7$ && $1/8$ && $p=3$: ~$\Phi$~~(associative)~~~~~ &\cr
&   &&  &&  &&  && $p=4$: ~$* \Phi$~~(coassociative) &\cr
&  && $Spin(7)$ && $8$ && $1/16$ && $p=4$: ~$\Psi$~~(Cayley)~~~~~~~~~ &\cr
\tablespace}\hrule}}}}
\centerline{ \hbox{{\bf Table 2:}{\it ~~ Examples of special
holonomy manifolds.}}}
} \vskip 0.5cm

All of these structures can be
characterized by the existence of a covariantly constant spinor,
\eqn\ccspinor{ \nabla \xi = 0 }
The existence of this $\xi$ guarantees that superstring
compactification on $X$ preserves some fraction (also listed in the
above table) of the original 32 supercharges, which is what makes such
manifolds useful in string theory.

Another characteristic property of special holonomy manifolds
is the existence of invariant forms, known as calibrations.
Using the covariantly constant spinor $\xi$
one can construct a $p$-form on $X$,
\eqn\pcalibr{ \om^{(p)} = \xi^{\dagger} \gamma_{i_1 \ldots i_p} \xi. }
By construction, such forms are covariantly constant and invariant
under ${\rm Hol} (X)$. They are non-trivial only for
special values of $p$:  see Table 2 for a list of the invariant forms on
manifolds of $SU(3)$, $G_2$, and $Spin(7)$ holonomy. These
invariant forms play an
important role in geometry and physics; in particular, they
can be used to characterize minimal (supersymmetric) cycles in
special holonomy manifolds.  Indeed, if $S \subset X$ is a minimal
submanifold of real dimension $p$, then its volume can be determined 
by integrating the invariant form $\omega^{(p)}$ over $S$,
\eqn\svolint{ \Vol (S) = \int_S \omega^{(p)}. }
Such a manifold $S$ is called calibrated, and the
form $\omega^{(p)}$ is called a calibration.  Notice the
simplification that has occurred here.  Ordinarily,
in order to compute the volume, $\Vol (S) = \int d^p x \sqrt{g}$,
we need to know the metric $g$; but the volume of a
calibrated submanifold $S$ is given by the simple formula \svolint\
which does not involve the explicit form of the metric.

This phenomenon is a prototype of various
situations in which the important geometric data can be characterized 
by differential forms rather than by a metric. 
This is essentially the same principle that was underlying the
constructions of Section 3, where we discussed form
theories of gravity in which the space-time geometry is encoded in
tensor forms and/or gauge fields.

To illustrate further the idea that forms can capture
the geometry, let us consider an example
which will play an important role below. Let $X$ be a manifold
with $G_2$ holonomy. The existence of
a $G_2$ holonomy metric is equivalent to the existence
of an associative 3-form, $\Phi$, which is closed and co-closed,
\eqn\phigtwo{\eqalign{
& d \Phi = 0 \cr
& d * \Phi = 0,
}}
and which can be written in terms of an orthonormal vielbein
$e^i$, $i=1, \ldots, 7$, as
\eqn\phioct{ \Phi = {1 \over 3!} \psi_{ijk} e^i \wedge e^j \wedge e^k. }
Here $\psi_{ijk}$ are the structure constants of the imaginary
octonions:  $\sigma_i \sigma_j = -\delta_{ij} + \psi_{ijk} \sigma_k$,
$\sigma_i \in \i (\IO)$.
Conversely, writing $\Phi$ locally in the form \phioct\ defines
a metric $g$ by the formula
\eqn\dsvia{
g = \sum_{i=1}^7 e^i \otimes e^i.
}
This $g$ can be written more explicitly by first defining
\eqn\bdef{ B_{jk} = - {1 \over 144}
\Phi_{ji_1 i_2} \Phi_{k j_3 j_4} \Phi_{j_5 j_6 j_7}
\epsilon^{j_1 \ldots j_7},
}
in terms of which the metric has a simple form,
\eqn\gviabb{ g_{ij} = \det (B)^{-1/9} B_{ij}. }
Evaluating the determinant of $B_{ij}$, we get $\det(g) =
\det(B)^{2/9}$, so \gviabb\ can be written in a more convenient form,
\eqn\ggss{ {\sqrt g}\,g_{jk} = - {1 \over 144} \Phi_{j i_1 i_2}
\Phi_{k i_3 i_4} \Phi_{i_5 i_6 i_7} \epsilon^{i_1 \ldots i_7}. }

Notice that even if the 3-form $\Phi$ does not obey \phigtwo, we
can still use \ggss\ to construct a metric on $X$ from $\Phi$,
as long as the 3-form $\Phi$ is non-degenerate in a suitable
sense.  (Of course, this metric will not have $G_2$ holonomy
unless \phigtwo\ is satisfied.)
This construction naturally leads us to the notion of stable forms,
which we now discuss.


\subsec{Stable Forms}

Following \Hitchin, in this section we review the construction of
action principles from $p$-forms.  It is natural to define the action
of a $p$-form $\rho$ on a manifold $X$ as a volume form $\phi (\rho)$
integrated over $X$.  One might think that such a $\phi(\rho)$ is hard
to construct, as one might have in mind the usual wedge product of
$\rho$ with itself, which gives a nonzero top-form only in rather
special cases.  In fact, this is not the only way to build a volume
form out of $p$-forms; as we will see, all the actions of interest for
us turn out to involve a volume element constructed in a rather
non-trivial way from the $p$-form.  For example, on a 7-manifold with
$G_2$ holonomy, the volume form cannot be constructed as a wedge
product of the associative 3-form $\Phi$ with itself; nevertheless,
one can define $\phi (\Phi)$ as a volume form for the metric \gviabb\
constructed from $\Phi$, as we will discuss in detail later.

The most general way to construct a volume form $\phi(\rho)$ from a
$p$-form $\rho$ is as follows: contract a number of
$\rho_{i_1,...,i_p}$'s with a number of epsilon tensors
$\epsilon^{i_1,...,i_n}$, to obtain some $W(\rho)$.  Suppose we use
$k$ epsilon tensors in $W$; then $W$ transforms as the $k$-th power of
a volume, and we can define $\phi(\rho)=W(\rho)^{1/k}$ which is a
volume form.  It is easy to see that $\phi(\rho)$ scales as an
$n/p$-th power of $\rho$ (if $W$ has $q$ factors of $\rho$ it will need
$k=pq/n$ factors of $\epsilon$).

Given such a volume form, one can define the
action to be the total volume,
\eqn\vxrho{ V (\rho) = \int_X \phi (\rho). }
This $V(\rho)$ is a homogeneous function of $\rho$ of degree ${n \over p}$:
\eqn\homognp{ V(\lambda \rho) = \lambda^{{n \over p}} V(\rho)
\quad\quad \lambda \in \IR. }

In this paper we will not be interested in arbitrary ways of putting
together $\rho$ and $\epsilon$ symbols to make $\phi(\rho)$.  Rather,
we will focus on cases in which there exists a notion of ``generic''
$p$-form.  In such cases the generic $p$-form $\rho$ defines an
interesting geometric structure ({\it e.g.} an almost complex structure or a
$G_2$ structure) even without imposing any additional constraints.
Hence arbitrary variations of $\rho$ can be thought of as variations
of this structure, and as we will see, critical points of $V(\rho)$
imply integrability conditions on these geometric structures.

The notion of genericity we have in mind is known as {\it stability},
as described in \Hit\ and reviewed below.  The requirement of
stability has drastically different consequences depending on the
dimension $n$ of the manifold and the degree $p$ of the form.  In most
cases, as we will see, there are no stable forms at all; but for
certain special values of $n$ and $p$, stable forms can exist.
Moreover, {\it all the calibrations in 6 and 7 dimensions that we
discussed earlier turn out to be stable forms.}  This deep
``coincidence'' makes the technology of stable forms a useful tool for
the study of special holonomy.  Nevertheless, possessing a stable form
is far less restrictive than the requirement of special holonomy,
needed for supersymmetry.

Let us now define the notion of stability precisely.  Write $V$ for
the tangent space at a point $x$, so the space of $p$-forms at $x$ is
$\Lambda^p V^*$.  Then a $p$-form $\rho$ is said to be stable at $x$
if $\rho(x)$ lies in an open orbit of the $GL(V)$ action on $\Lambda^p
V^*$.  We call $\rho$ a stable form if $\rho$ is stable at every
point.  In other words, {\it $\rho$ is stable if all forms in a
neighborhood of $\rho$ are equivalent to $\rho$ by a local $GL(n)$
action.}

Some special cases of stability are easy to understand.  For example,
there are no stable $0$-forms, because under coordinate transformations the
value of a function does not change.  On the other hand, any nonzero
$n$-form is stable, because by a linear transformation one can always 
map any volume form to any other one.  Similarly, any nonzero $1$-form
or $(n-1)$-form is stable.

A less trivial case of stability is that of a $2$-form, on a manifold of even 
dimension.  In this case, viewing the $2$-form as an 
antisymmetric matrix, stability just means that the determinant
is nonzero; namely, this is the usual characterization of a
presymplectic form, and any such form can be mapped to any other
by a linear transformation, so they are indeed stable.
Given such a stable form we can now construct its associated
volume form:  namely, we write $\phi(\omega)=\omega^{n/2}$.  
Note that the stability of $\omega$ 
is equivalent to $\phi(\omega) \ne 0$.

To understand the geometric structures defined by stable forms,
it is useful to study the subgroup of $GL(n)$ which fixes
them.  In the case of a stable $2$-form in even dimension this stabilizer is $Sp(n)$,
corresponding to the fact that the $2$-form defines a presymplectic structure.
More generally, given a stable $p$-form, we can easily compute
the dimension of the stabilizer:  it is simply the
dimension of $GL(n)$ minus the dimension of the space of
$p$-forms.  In the case $p=2$, this counting gives $n^2-{n(n-1)\over 2}=
n(n+1)/2$, as expected.

Next we consider the case $p=3$. The dimension of the space of
3-forms is
\eqn\dimthreef{{\rm dim}\ \Lambda^3 V^* = n(n-1)(n-2)/6.}
Already at this stage we see that there cannot be any stable 3-forms
for large $n$, because $\dim GL(n) = n^2$ has a slower growth than
$n^3/6$, so that $GL(n)$ cannot act locally transitively on the space
of 3-forms.  However, for $p=3$ and small enough $n$ the stability
condition can be met.  We have already discussed the cases $n=3,4$,
and stable 3-forms also exist for $n=6,7,8$.  These special cases lead
to interesting geometric structures; for example, consider the case
$p=3, n=7$.  Here the dimension counting gives
\eqn\gtwocount{\eqalign{
\dim GL(V) = n^2 =\ &49, \cr
\dim \Lambda^p V^* = {n! \over p! (n-p)!}  =\ &35, \cr
&14 = \dim G_2.~~~
}}
Indeed, the stabilizer of the 3-form in this case is $G_2$, so a stable 3-form
in 7 dimensions defines a $G_2$ structure.

As we just discussed for $p=3$,
generically $\dim \Lambda^p V^*$ is much larger than
$\dim GL(V) = n^2$.  Hence stable forms
exist only for special values of $n$ and $p$ \Hit.
The cases of interest for us in this paper are
$n=7$ with $p=3,4$ and $n=6$ with $p=3,4$.
We now turn to the construction of volume functionals
from stable forms in these cases.

\subsec{3-Form and 4-Form Actions in $6D$}

We begin with the 6-dimensional case.  In this case Hitchin
constructed two action functionals $V_H(\rho)$ and $V_S(\sigma)$,
depending respectively on a $3$-form $\rho$ and $4$-form $\sigma$.
When extremized, $V_H$ and $V_S$ yield respectively holomorphic and
symplectic structures on $M$.  In this section we introduce these
action functionals and describe some of their properties.

Let us first construct $V_S(\sigma)$.  Suppose $\sigma$ is a stable
4-form; the stability condition in this case means there exists $k$
such that $\sigma = \pm \half k \wedge k$.  We consider the $+$ case here.
Interpreting this $k$ as a
candidate symplectic form, $V_S(\sigma)$ is defined to be the
symplectic volume of $M$:
\eqn\defVsigma{
V_S(\sigma) = {1 \over 6} \int_M k \wedge k \wedge k.
}
This action can also be written directly in terms of $\sigma$:
\eqn\vsigma{\eqalign{
V_S(\sigma)
& = {1 \over 6} \int_M \sigma^{3/2} = \cr
& = \int_M \sqrt{ {\textstyle{1 \over 384}}
\sigma_{a_1 a_2 b_1 b_2} \sigma_{a_3 a_4 b_3 b_4} \sigma_{a_5 a_6 b_5 b_6}
\epsilon^{a_1 a_2 a_3 a_4 a_5 a_6} \epsilon^{b_1 b_2 b_3 b_4 b_5 b_6}
}, }}
where $\epsilon^{a_1 \ldots a_6}$ is the Levi-Civita tensor in six
dimensions.  As discussed before, the need to take a square
root arises because to define a volume form 
we need to have exactly one net $\epsilon$ tensor.

We will be considering $V_S(\sigma)$ as the effective action of
a gravity theory in six dimensions.  We treat $\sigma$ as a
4-form field strength for a 3-form gauge field $\gamma$:  in other words,
we hold the cohomology class of $\sigma$ fixed,
\eqn\varysigma{\eqalign{
& [\sigma] \in H^4 (M,\R) ~~~~~~{\rm fixed},
~i.e. \cr & \sigma = \sigma_0 + d \gamma, }}
where $d \sigma_0 = 0$.  Now we want to find the classical solutions,
i.e. critical points of $V_S(\sigma)$.  Write
\eqn\vsigmaagain{
V_S(\sigma) = {1 \over 3} \int_M \sigma \wedge k = {1 \over 3} \int_M (\sigma_0 + d \gamma) \wedge k.
}
Varying $\gamma$ then gives a term
\eqn\firstvar{
\delta V_S = {1 \over 3} \int_M d (\delta \gamma) \wedge k = - {1 \over 3} \int_M \delta \gamma \wedge dk.
}  
This is not the whole variation of $V_S$, because $k$ also depends on $\sigma$; but it turns out that the
extra term from the variation of $k$ is just $1/2$ of \firstvar.  This is a consequence of the fact \homognp\ 
that $V_S(\sigma)$ is homogeneous as a function of $\sigma$.  Altogether, the
condition that $V_S(\sigma)$ is extremal under variations of $\gamma$ is simply
\eqn\conditionk{ dk = 0. }
Hence the classical solutions of the gravity theory based on 
$V_S(\sigma)$ give symplectic structures on $M$.

Having discussed $V_S(\sigma)$, we now turn to $V_H(\rho)$.  
Suppose $\rho$ is a stable 3-form.  Provided that $\rho$ is ``positive'' 
in a sense to be defined below, it is fixed by a subgroup of $GL(6,\R)$ 
isomorphic to (two copies of) $SL(3,\C)$;
this $\rho$ therefore determines a reduction of $GL(6)$ to $SL(3,\C)$,
which is the same as an almost complex structure on $M$.  
More concretely, we can find three complex 1-forms $\zeta_i$ for which
\eqn\rhoform{
\rho = \half (\zeta_1 \wedge \zeta_2 \wedge \zeta_3 + \bar{\zeta_1} \wedge \bar{\zeta_2}
\wedge \bar{\zeta_3}),
}
and these $\zeta_i$ determine the almost complex structure.  If locally there exist complex
coordinates such that $dz_i = \zeta_i$, then the almost complex structure is integrable 
(it defines an actual complex structure.)
Whether it is integrable or not, we can construct a $(3,0)$ form on $M$, namely
\eqn\omegasmall{
\Omega = \zeta_1 \wedge \zeta_2 \wedge \zeta_3.
}
This $\Omega$ can also be written
\eqn\defOmega{ 
\Omega = \rho + i \hat{\rho}(\rho),
}
where $\hat{\rho}$ is defined as
\eqn\defhrho{
\hat\rho = - \ihalf (\zeta_1 \wedge \zeta_2 \wedge \zeta_3 - \bar{\zeta_1} \wedge \bar{\zeta_2}
\wedge \bar{\zeta_3}).
}
Through \rhoform, \omegasmall\ and \defhrho, we can regard $\Omega$ and $\hat\rho$
as functions of $\rho$.  The integrability condition is equivalent to the
requirement that $d \Omega = 0$.

So far we have explained how a positive stable 3-form $\rho$ determines an almost
complex structure and a holomorphic 3-form $\Omega$. 
Now $V_H(\rho)$ is defined to be the holomorphic volume:
\eqn\defVrho{ V_H(\rho) = - {i \over 4} \int_M \Omega \wedge \bar{\Omega}
= {1 \over 2} \int_M \hat{\rho}(\rho) \wedge
\rho. }
More concretely,
using results from \Hitchin, this can be written
\eqn\vrho{ V_H(\rho)
= \int_M \sqrt{- {\textstyle{1 \over 6}} {K_a}^b {K_b}^a }, }
where\foot{Having written this formula we can now explain the positivity
condition on $\rho$ to which we alluded earlier:  the square root
which appears in \vrho\ should be real.}
\eqn\krhorho{
{K_a}^b := {1 \over 12} \rho_{a_1 a_2 a_3} \rho_{a_4 a_5 a}
\epsilon^{a_1 a_2 a_3 a_4 a_5 b}. }
As we did with $V_S$, we want to regard $V_H$ as the effective action of some
gravity theory in which $\rho$ is treated as a field strength.  So
we start with a closed stable 3-form $\rho_0$ and allow it
to vary in a fixed cohomology class,
\eqn\varyrho{ \rho = \rho_0 + d\beta. }
Then varying $\beta$, we obtain two terms, one from the variation
of $\rho$ and one from the variation of $\hat\rho(\rho)$.  As
in the case of $V_S$, the homogeneity of $V_H(\rho)$ implies that
these two terms are equal, and they combine to give
\eqn\variationrho{
\delta V_H = \int_M d(\delta \gamma) \wedge \hat\rho = - \int_M \delta \gamma \wedge d \hat\rho.
}
Hence the equation of motion is 
$$
d \hat \rho=0.
$$
{}From \varyrho\ we also have $d \rho=0$.  So altogether on-shell we have 
$d \Omega = 0$, which is the condition for integrability of the
almost complex structure, as explained above.  In
this sense, $V_H(\rho)$ is an action which generates complex
structures together with holomorphic three-forms.

Finally, let us make one more observation about the functionals $V_H$
and $V_S$.  So far we have discussed them separately, but since they
both exist on a 6-manifold $M$, it is natural to ask whether the
structures they define are compatible with one another.  Specifically,
we would like to interpret $k$ as the K\"ahler form on $M$, in the
complex structure determined by $\Omega$.  For this interpretation to
be consistent, there are two conditions which must be satisfied:
\eqn\compata{ k \wedge \rho = 0, }
and
\eqn\compatb{ 2V_S(\sigma) = V_H(\rho). }
The condition \compata\ expresses the requirement that $k$ is of
type $(1,1)$ in the complex structure determined by $\Omega$, 
while \compatb\ is the equality of the volume forms determined
independently by the holomorphic and symplectic structures.
Requiring \compata--\compatb, $\Omega$ and $k$ together give 
an $SU(3)$ structure on $M$; if in addition $d \Omega = 0$, $dk = 0$, then $M$ is
Calabi-Yau, with $\Omega$ as holomorphic 3-form and $k$
as \kahler\ form.
When we discuss the Hamiltonian quantization of topological M-theory
in Section 7, we will see one way in which these constraints could arise naturally.

\subsec{3-Form and 4-Form Actions in $7D$}

Now let us discuss the 7-dimensional case.  We will construct
two functionals $V_7(\Phi)$, $V_7(G)$ depending on a 3-form
or 4-form respectively, both of which generate $G_2$ holonomy
metrics on a 7-manifold $X$.

The two cases are very similar to one another; we begin with the 
3-form case.  A stable
3-form $\Phi \in \Omega^3 (X,\R)$ determines a $G_2$ structure on
$X$, because $G_2$ is the subgroup of
$GL(7,\R)$ fixing $\Phi$ at each point, as we explained in Section 4.2.
There we gave the explicit expression for the metric 
$g$ in terms of the 3-form $\Phi$:
\eqn\gbb{ g_{jk} = B_{jk} \det(B)^{-1/9}, }
where from \bdef,
\eqn\bphidef{ B_{jk} = -{1 \over 144} \Phi_{j i_1 i_2} \Phi_{k i_3
i_4} \Phi_{i_5 i_6 i_7} \epsilon^{i_1 \ldots i_7}. }

We can thus introduce a volume functional, $V_7(\Phi)$, which is simply 
the volume of $X$ as determined by $g$:
\eqn\Vseven{ V_7 (\Phi) = \int_X \sqrt{g_{\Phi}} = \int_X \big(
\det B \big)^{1/9}, }
where $B$ is the symmetric tensor defined in \bphidef. 

In order to identify the critical points of the action functional
\Vseven, it is convenient to rewrite it slightly.  For this we 
use the fact that since $\Phi$ determines the metric, it also
determines any quantity which could be derived from the metric; in
particular it determines a Hodge $*$-operator, which we write
$*_{\Phi}$.  Using this operator we can rewrite \Vseven\ as
\eqn\VPhi{ V_7 (\Phi) = \int_X \Phi \wedge *_{\Phi} \Phi. }
As we did in the 6-dimensional cases, we regard $\Phi$ as a field
strength for a 2-form gauge potential; in other words,
we assume $\Phi$ is closed and vary it in a fixed cohomology class:
\eqn\varyphi{\eqalign{ & [\Phi] \in H^3 (X,\R) ~~~~~~{\rm fixed},
~i.e. \cr & \Phi = \Phi_0 + d B, }}
with $d \Phi_0 = 0$, and $B$ an arbitrary real 2-form on $X$.
Using the homogeneity property \homognp\ of the volume functional \VPhi, we find
\eqn\varyphi{ {\delta V_7 (\Phi) \over \delta \Phi} = {7 \over 3} *_{\Phi}
\Phi. }
Hence critical points of
$V_7 (\Phi)$ in a fixed cohomology class give
3-forms which are closed and co-closed,
\eqn\ddphi{\eqalign{ & d \Phi = 0, \cr & d *_\Phi \Phi = 0. }}
These are precisely the conditions under which $\Phi$ 
is the associative 3-form for a $G_2$ holonomy
metric on $X$.

So far we have discussed stable 3-forms, but the $G_2$ holonomy 
condition can also be obtained from a dual action based on a
stable 4-form $G$,
\eqn\VG{ V_7(G) = \int_X G \wedge *_G G. }
It is this $V_7(G)$ which we propose to identify as the
effective action of the 7-dimensional topological M-theory.  As in
the previous cases, we vary the 4-form $G$ in a fixed cohomology
class:
\eqn\varyg{\eqalign{ & [G] \in H^4 (X,\R) ~~~~~~{\rm fixed}, ~i.e.
\cr & G = G_0 + d \Gamma, }}
where $\Gamma$ is an arbitrary real 3-form on $X$, and $G_0$ is
closed, $d G_0 = 0$. The condition that \VG\ is
extremal under variations of $\Gamma$ is then simply
\eqn\ddg{\eqalign{ & d G = 0, \cr & d *_G G = 0, }}
which is again the condition \ddphi\ that $X$ has $G_2$
holonomy, now written in terms of the coassociative 4-form $G =
*_{\Phi} \Phi$.  Just as with $\Phi$, one can reconstruct
the $G_2$ holonomy metric from $G$, using the expression
of $G$ in terms of an orthonormal vielbein,
\eqn\vielbeing{ G = e^{7346} - e^{7126} + e^{7135} - 
e^{7425} + e^{1342} + e^{5623} + e^{5641}. }

The 4-form action \VG\ can also be written in a slightly different form.
One introduces a fixed basis of the space $\wedge^2 V$ of bivectors in 7 dimensions:
$e_a^{ij}=-e_a^{ji}$.  Here $i,j=1,\ldots,7$ and $a=1,\ldots,21$, since
the space of bivectors is 21-dimensional. 
Then define the matrix
$Q_{ab}$ by
\eqn\Qmatr{
Q_{ab} = {1\over 2} e_a^{ij} e_b^{kl} G_{ijkl}.
}
The action for $G$ can then be written as
\eqn\SofQ{
V_7(G) =  \int_X (\det Q)^{1\over 12}.
}
Note that since $Q$ is a matrix of rank 21, this action is indeed
homogeneous of degree ${21\over 12} = {7\over 4}$ in $G$. It is a
tempting thought that this action could be interpreted as a membrane
version of Born-Infeld obtained by integrating out open (topological)
membranes, since the exponent ${1\over 12}$ reminds one of a stringy
one-loop determinant.

\subsec{Hamiltonian Flow}

Now we shift gears to discuss a bridge between the 
$SU(3)$ structures and $G_2$ holonomy metrics considered
in the last two subsections:  we will describe a flow
which constructs $G_2$ holonomy metrics from the $SU(3)$
structures which appeared there.  
This flow is essentially a Hamiltonian version
of the Lagrangian field theories described in Section 4.4.

Suppose given a 6-manifold $M$, 
with stable forms $\rho \in \Omega^3 (M,\R)$ and $\sigma \in
\Omega^4 (M,\R)$.  As we discussed above, if $\rho$ and $\sigma$
satisfy the compatibility conditions \compata\ and \compatb, they
define an $SU(3)$ structure on $M$ and a corresponding metric.  If
$\rho$ and $\sigma$ are also both closed, one can extend the metric on
$M$ uniquely to a $G_2$ holonomy metric on $X = M \times (a,b)$ for
some interval $(a,b)$.  Hitchin gave an elegant construction of this
metric \Hitchin: one takes the given $\rho$ and $\sigma$ as ``initial
data'' on $M \times \{t_0\}$ and then lets $\rho$ and $\sigma$ flow
according to
\eqn\hflow{\eqalign{
& {\p \rho \over \p t} = d k, \cr
{\p \sigma \over \p t}
& =  k \wedge {\p k \over \p t} = - d \hat \rho.
}}
Here, as usual, $\sigma = \half k \wedge k$,
and $t$ is the ``time'' direction normal to $M$.

\ifig\slagfig{A $G_2$ structure on the local geometry $X = M
\times (a,b)$ can be viewed as a Hamiltonian flow.}
{\epsfxsize2.5in\epsfbox{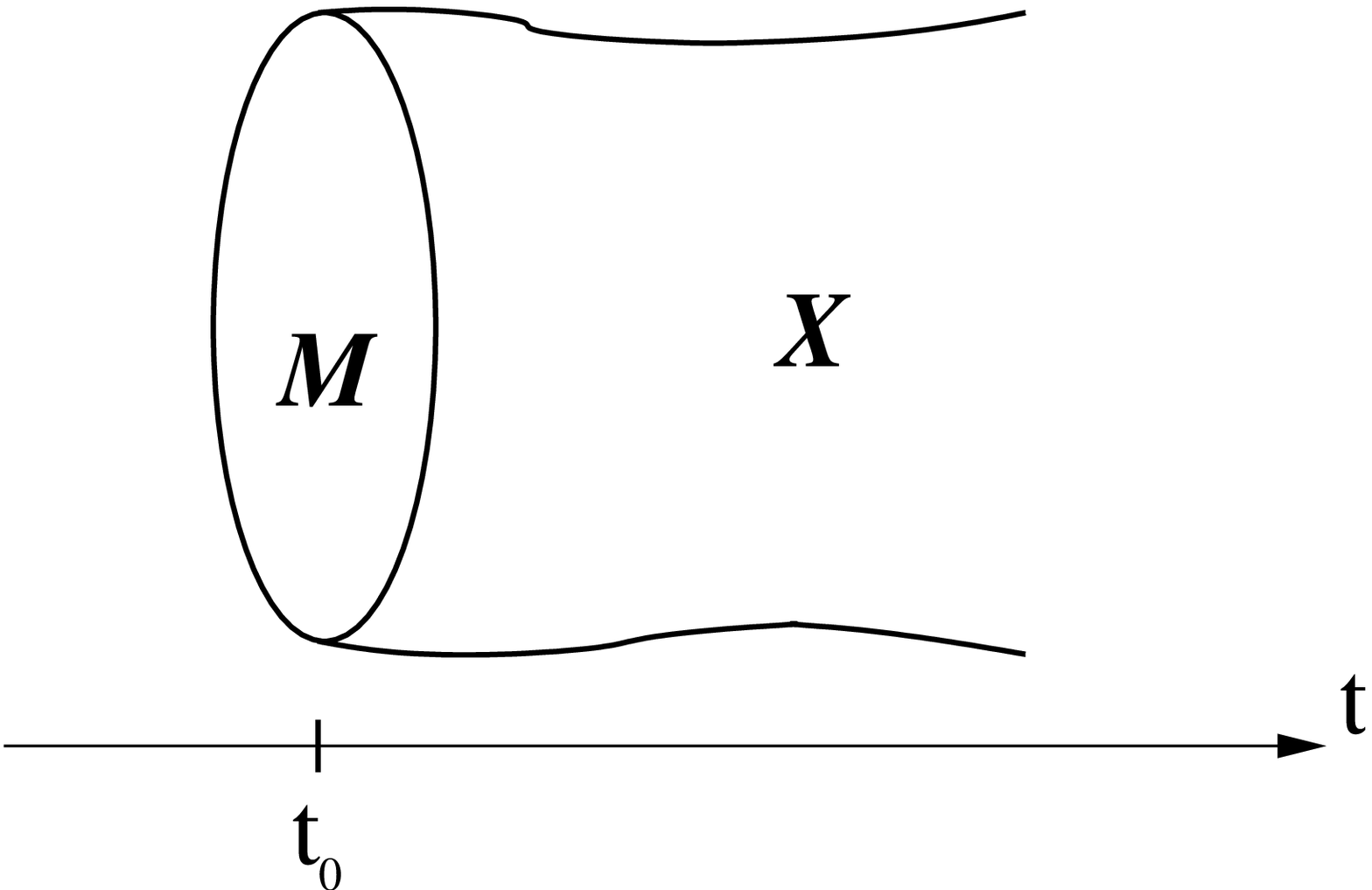}}

The evolution equations \hflow\ are equivalent to the $G_2$
holonomy conditions \ddphi\ for the 3-form
$$\Phi = \rho (t) + k (t) \wedge dt.$$
Moreover, \hflow\ can be interpreted as Hamiltonian flow
equations.  Namely, one considers the variations of $\sigma$ and
$\rho$ as spanning a phase space $\Omega^4_{exact}(M,\R) \times \Omega^3_{exact}(M,\R)$; writing
$\delta \sigma = d \beta$ and $\delta \rho = d \alpha$, the symplectic
pairing on the phase space is
\eqn\pairing{
\langle \delta \sigma, \delta \rho \rangle = \int \alpha \wedge d
\beta = - \int \beta \wedge d \alpha.
}
Then \hflow\ are precisely the Hamiltonian flow equations with
respect to
\eqn\hamilt{H = 2 V_S(\sigma) - V_H(\rho),}
where $V_H(\rho)$ and $V_S(\sigma)$ are the volume functionals
\defVsigma\ and \defVrho\ which we used to obtain $SU(3)$ structures
in 6 dimensions.

\newsec{Relating Hitchin's Functionals in $6D$ to Topological Strings}

In the last section we introduced two functionals $V_H(\rho)$,
$V_S(\sigma)$ which, when extremized, generate respectively a
symplectic form $k$ and a closed holomorphic $(3,0)$ form $\Omega$ on
a 6-manifold $M$.  This is reminiscent of the topological A and B
models, and one might wonder whether there is some relation.  In this
section we point out that such a relation does exist.  Our arguments
will be rigorous only at the classical level, but they suggest a
natural extension to the quantum theories, which we will describe.
One partcularly interesting feature will emerge: namely, $V_H(\rho)$
turns out to be equivalent not to the B model itself but to a
combination of the B and $\bar{\rm B}$ models.

\subsec{Hitchin's $V_S$ as the A model}

We begin by discussing
a relation between Hitchin's action functional \defVsigma,
\eqn\saction{ V_S (\sigma) = \int_M \sigma^{3/2} }
based on the closed 4-form $\sigma$, and the A model on $M$.  As we
discussed in Section 4.3, the solutions to the classical equations of
motion coming from $V_S(\sigma)$ involve K\"ahler geometries, which
are also the classical solutions of the A model K\"ahler gravity.  In
fact, more is true: the classical actions in both cases compute the
volume of $M$.  So at least at a superficial classical level, the two
theories are equivalent.  Moreover, we can argue that the small 
fluctuations in the two
theories can be identified with one another.  Namely, recall that in
Hitchin's theory we write $\sigma =\sigma_0+d\gamma$; then the action
at quadratic order for the fluctuation $\gamma$ includes $\int d\gamma
\wedge {d^c}^\dagger \gamma$, which nicely matches the action for the
quadratic fluctuations in the \kahler\ gravity theory.  So one would
expect that the two should be identified.

Here we would like to take one more step in connecting the two theories:
specifically,
it has been recently argued \INOV\ that the A model can be reformulated in
terms of a topologically twisted $U(1)$ gauge theory on $M$, whose bosonic
action contains the observables
\eqn\foamact{ S = {g_s \over 3} \int_M F \wedge F \wedge F
+ \int_M k_0 \wedge F \wedge F.}
The partition function in this theory is a function of the fixed class
$k_0$.  The path integral can be defined as a sum over a gravitational
``quantum foam'' \INOV, {\it i.e.} over K\"ahler geometries with
quantized K\"ahler class,\foot{We choose our normalization so that $F$
is integrally quantized: $\int_C F \in \Z$ for any closed 2-cycle $C
\subset M$.}
\eqn\kclass{ k = k_0 + g_s F, }
or, equivalently, as a sum over ideal sheaves \MNOP.

We claim that in the weak coupling ($g_s \to 0$) limit, the theory
based on the action \foamact\ is equivalent to the ``gravity theory''
based on Hitchin's action \saction.  Moreover, we show that fixing the
BRST symmetries of the Hitchin action naturally leads to the
description of the A model as a topologically twisted supersymmetric
$U(1)$ gauge theory.

In order to show this, we begin with the action
\eqn\maction{S = \a \int_M \tilde F \wedge \tilde F \wedge \tilde F
- \b \int_M \sigma \wedge \tilde F , }
where $\a$ and $\b$ are some coefficients (which will be related to
$g_s$ below), $\tilde F$ is a 2-form on $M$, and
$\sigma$ is a 4-form that varies in a fixed cohomology class,
\eqn\svary{\eqalign{
& [\sigma] \in H^4 (M) ~~~~~~{\rm fixed}, ~i.e. \cr
& \sigma = \sigma_0 + d \gamma.
}}
At this point we do not make any assumptions about
the 2-form $\tilde F$; in particular, it need not be closed
or co-closed.

First, let us integrate out the 2-form $\tilde F$
in the action \maction.
The equation of motion for $\tilde F$ has the form
\eqn\feom{ 3\a \tilde F \wedge \tilde F - \b \sigma = 0.}
This equation implies that the 2-form $\tilde F$ is a "square
root" of $\sigma$, {\it i.e.} $\sigma$ is a stable 4-form.
Substituting $\tilde F$ back into the action \maction, we obtain
precisely Hitchin's action \saction, with the remaining path
integral over a stable, closed 4-form $\sigma$. It is important to
stress here that, since the action \maction\ is cubic in $\tilde
F$, the relation to Hitchin's action \saction\ holds only in
the semi-classical limit. We return to this issue below, and show
that this is precisely the limit $g_s \to 0$.

Similarly, starting with the action \maction\ and integrating out
$\sigma$ one can obtain the $U(1)$ gauge theory \foamact.  In order to
see this, one has to eliminate $\sigma$ through its equations of
motion, and then make a simple field redefinition.  The equations of
motion for $\sigma$ are very simple.  Since the dynamical variable
$\gamma$ appears as a Lagrange multiplier in \maction, it leads to the
constraint
\eqn\fclosed{ d \tilde F=0, }
which means that the 2-form $\tilde F$ is closed and, therefore,
can be interpreted as a curvature on a line bundle.
The resulting action for $\tilde F$ is
\eqn\fact{S = \a \int_M \tilde F \wedge \tilde F \wedge \tilde F
- \b \int_M \sigma_0 \wedge \tilde F.}
In order to bring this action to the familiar form \foamact, it
remains to do a simple change of variables. We introduce
\eqn\fchange{F = \tilde F - \xi k_0,}
where $\xi$ is a parameter and $k_0$ is the background K\"ahler
form, such that $\sigma_0 = k_0 \wedge k_0$. Substituting
\fchange\ into \fact, we get (up to a
constant term) the action
\eqn\ffact{
S = \a \int_M F \wedge F \wedge F + 3 \xi \a \int_M F \wedge F \wedge k_0
+ \int_M \Big( 3 \xi^2 \a k_0 \wedge k_0 \wedge F - \b \sigma_0 \wedge F \Big).
}
Comparing \ffact\ with \foamact\ determines the
parameters $\a$, $\b$, and $\xi$:
\eqn\abxifix{\eqalign{
& \a = {g_s \over 3}, \cr
& \xi = {1 \over g_s}, \cr
& \b = {1 \over g_s}.
}}
With this choice of parameters, we find complete agreement
between \ffact\ and the $U(1)$ gauge theory action \foamact,
including the numerical coefficients and the relation
between the K\"ahler form $k$ and the field $F$. Indeed,
substituting $\xi=1/g_s$ into \fchange, we get
\eqn\kquant{ \delta k = g_s F,}
which is precisely the required quantization condition \kclass.

Summarizing, we find that \maction\ is equivalent to the gauge
theory action \foamact\ and, in the semi-classical limit, is also
equal to Hitchin's action \saction. In order to see when the
semi-classical approximation is valid, it is convenient to write
both terms in the action \maction\ with the same overall
coefficient $1/\hbar$. To achieve this, we rescale
\eqn\trescale{
\tilde F \to \gamma \tilde F,
}
and set the coefficients in the two terms to be equal:
\eqn\hbarlim{ {1 \over \hbar} = \a \g^3 = \g. }
In particular, the latter equality implies $ \a = {1 \over \g^2}$.
{}From the relations \abxifix\ and \hbarlim\ it follows that the
semiclassical limit, $\hbar \to 0$, corresponds to the limit $g_s
\to 0$. Hence we conclude that the gauge theory action \foamact\
is equivalent to Hitchin's action \saction\ precisely in the
weak coupling limit.

\bigskip
\noindent {\it BRST Symmetries and Gauge Fixing}
\medskip

As noted before, we really want to connect Hitchin's theory
to the {\it topologically twisted} version of the supersymmetric 
$U(1)$ gauge theory.  In order to do this
let us describe the BRST symmetries of the theory \maction, which,
as we just established, is equivalent to the $U(1)$ gauge theory
\foamact. First, notice that the partition sum over the quantum
foam can be viewed as a vacuum expectation value,
\eqn\oovev{ 
\langle \exp \Big( {g_s \over 3} \int \CO_1 + \int \CO_2 \Big)  \rangle,
}
in the topological $U(1)$ gauge theory on $M$,
where $\CO_i$ are the topological observables:
\eqn\observables{\eqalign{
\CO_1  & = F \wedge F \wedge F, \cr
\CO_2  & = k_0  \wedge F \wedge F, \cr
& \vdots
}}
Following \BaulieuS, one can reconstruct the action of this
topological 6-dimensional theory by studying the BRST symmetries
that preserve \oovev--\observables. Writing (locally) $F$ as a
curvature of a gauge connection $A$,
\eqn\fda{ F = dA, }
it is easy to see that \oovev--\observables\ are invariant under
the usual gauge transformations
\eqn\agauge{ \delta A = d \epsilon_0,}
as well as under more general transformations
\eqn\dela{ \delta A = \epsilon_1}
where the infinitesimal parameter $\epsilon_1$ is a 1-form on $M$.
The gauge fixing of the latter symmetry leads to a 1-form ghost field
$\psi$. Since $\epsilon_1$ itself has a gauge symmetry, $\epsilon_1
\sim \epsilon_1 + d \lambda$, one also has to introduce a commuting
0-form $\phi$ associated with this symmetry.  Hence, already at this
stage we see that the $6D$ topological theory in question should contain
a gauge field and a scalar. The only such theory is a maximally
supersymmetric topological gauge theory in six dimensions, {\it i.e.}
a theory with $\CN_T = 2$ topological supersymmetry. Equivalently, it
is a theory with 16 real fermions, which can be identified with
holomorphic $(p,0)$-forms on $M$. Remember that on a K\"ahler
manifold
\eqn\spinvshol{S(M) \cong \Omega^{0,*} (M). }
The complete BRST multiplet in this theory looks like:
\eqn\sixspectr{\eqalign{
{\rm Bosons:} \quad\quad\quad 1-{\rm form} & \quad A \cr
{\rm cplx. ~scalar} & \quad \phi \cr
(3,0)-{\rm form} & \quad \varphi \cr
{\rm Fermions:} \quad\quad\quad \quad\quad\quad\quad
& \quad \psi^{p,0} ~,~\psi^{0,p} \quad\quad p=0,1,2,3  \cr
}}
Under the action of the BRST operator $s$, these fields transform
as \refs{\BKS,\HPark}:
\eqn\brstbks{\eqalign{
& s \varphi^{0,3} = 0 \quad\quad\quad\quad\quad~ s \varphi^{3,0} =
\psi^{3,0} \cr
& s A^{0,1} = \psi^{0,1} \quad\quad\quad\quad  s A^{1,0} =0 \cr
& s \psi^{0,1} = 0 \quad\quad\quad\quad\quad~  s \psi^{1,0} = -
\p_A \phi \cr
& s \psi^{0,0} = ( k \cdot F^{1,1} ) \quad\quad  s \psi^{2,0} =
F^{2,0}
 }}
This $\CN_T=2$ 6-dimensional topological $U(1)$ gauge theory has
been extensively studied in the literature, see {\it e.g.}
\refs{\BKS,\HPark,\DT,\Acharya,\BlauT}. A reduction of this theory
on a K\"ahler 4-manifold $M^4 \subset M$ leads to the $\CN_T=4$
topological gauge theory studied in \VW.

Finally, we identify the symmetry of Hitchin's action
\saction\ that corresponds to the BRST symmetry \dela. In order to
do this, we need to find how this symmetry acts on the 4-form
field $\sigma$. Since in the $g_s \to 0$ limit the field $F$ is
identified with the variation of the K\"aher form \kquant\ it
follows that
\eqn\delk{ \delta k = d \epsilon_1, }
where $\sigma = k \wedge k$. It is easy to check that
Hitchin's action \saction\ is indeed invariant under this
symmetry,
\eqn\invsymmh{
\delta S_H = {3 \over 2} \int_M k \wedge \delta \sigma = 3 \int_M
k \wedge k \wedge \delta k = 3 \int_M \sigma \wedge d\epsilon_1 =
0.
}

We have thus recovered the topologically twisted $U(1)$ theory
which was conjectured in \INOV\ to be equivalent to the quantum
foam description of the A model.


\subsec{Hitchin's $V_H$ as the B model}

Now we want to discuss the relation between Hitchin's ``holomorphic
volume'' functional $V_H(\rho)$ and the B model (see also the recent
work \SGerasimov, which proposes a relation similar to what we will
propose below.)  Classically, there is an obvious connection between
the two, since solving the equations of motion of either one gives a
closed holomorphic 3-form $\Omega$.  What about quantum mechanically?
In order to understand this question we must first recall a subtle
feature of the B model partition function.

Consider the B model on a Calabi-Yau 3-fold $M$.  This model is
obtained by topological twisting of the physical theory with a fixed
``background'' complex structure, determined by a holomorphic 3-form
$\Omega$.  The topological observables in this model are the marginal
operators $\phi_i$ representing infinitesimal deformations of the
complex structure, where $i=1, \dots, h_{2,1}$.  The partition
function $Z_B$ was defined in \BCOV\ to be the generating functional
of correlations of marginal operators: namely, introducing $h_{2,1}$
variables $x^i$, $Z_B(x,g_s,\Omega_0)$ obeys
\eqn\defz{
D_{i_1} \cdots D_{i_k} Z_B(x,g_s) |_{x=0} = \langle
\phi_{i_1} \cdots \phi_{i_k} \rangle_{\Omega_0}.}
More intrinsically, we can think of $x$ as labeling an infinitesimal
deformation, so that for fixed $\Omega_0$, $Z_B(x,g_s,\Omega_0)$ is a
function on the holomorphic tangent space $T_{\Omega_0} \CM$ to the
moduli space $\CM$ of complex structures.  By construction this $Z_B$
is holomorphic in its dependence on $x$.  But one gets one such
function of $x$ for every $\Omega_0$, corresponding to all the
different tangent spaces to $\CM$, and one can ask how these different
functions are related.  This question was answered in \BCOV, where the
effect of an infinitesimal change in $\Omega_0$ was found to be given
by a ``holomorphic anomaly equation.''

This $\Omega_0$ dependence of $Z_B$ was later reinterpreted in
\Wittenhol\ as the wavefunction property.  To understand what this
means, it is convenient to combine $g_s$ and $x$ into a ``large phase
space'' of dimension $h_{2,1}+1$.  Changing $g_s$ is equivalent to an
overall rescaling of $\Omega_0$ (which does not change the complex
structure on $X$).  So for fixed $\Omega_0$, we can consider $Z_B$ as
a holomorphic function on $H^{3,0}(X,\C) \oplus H^{2,1}(X,\C)$.
Equivalently, $Z_B$ is a function on the ``phase space'' $H^3(X,\R)$,
which depends only on the complex combination of periods
\eqn\okcombo{
x_I = F_I - \tau_{IJ}(\Omega_0) X^J,
}
and not on the conjugate combination
\eqn\conjcombo{
\bar{x}_I = F_I - \overline{\tau}_{IJ}(\Omega_0) X^J.
} 
This is similar to the idea of a wavefunction which depends only on
$q$ but not on its conjugate variable $p$; indeed, $x_I$ and
$\bar{x}_I$ are coordinates on $H^{3,0} \oplus H^{2,1}$ and on
$H^{1,2} \oplus H^{0,3}$ respectively, and they are indeed conjugate
with respect to the standard symplectic structure on $H^3(X,\R)$.
Note that since $\tau$ depends on $\Omega_0$, changing $\Omega_0$
changes the symplectic coordinate system.

Now, if one is given a wavefunction $\psi(q)$ as a function of $q$ and
one wants to express it as a function of $p$, there is a simple
procedure for doing so: just take the Fourier transform.  In fact,
more generally, given $\psi(q)$ one can construct various different
representations of the state, {\it e.g.} $\psi(p)$, $\psi(p+q)$,
$\psi(p+iq)$ and so on.  Each such representation corresponds to a
different choice of symplectic coordinates inside the $(p,q)$ phase
space, and each can be obtained from $\psi(q)$ by an appropriate
generalized Fourier transform.  In \Wittenhol\ it was shown that the
$\Omega_0$ dependence of $Z_B$ can be understood in exactly this way:
starting from $Z_B(x,\Omega_0)$ one can obtain any other
$Z_B(x,\Omega'_0)$ by taking an appropriate Fourier transform!  In
this sense $Z_B$ is a wavefunction obtained by quantization of the
symplectic phase space $H^3(X,\R)$, which has various different
representations depending on one's choice of symplectic coordinates
for $H^3(X,\R)$.

Now what about Hitchin's gravity theory?  Consider the partition
function $Z_H([\rho])$ of Hitchin's 6-dimensional gravity theory,
formally written
\eqn\zhitchin{
Z_H([\rho]) = \int D \beta \exp( V_H(\rho + d\beta) ).  } 
We do not expect that the formal expression \zhitchin\ really captures
the whole quantum theory, but the statement that $Z_H$ depends on a
class $[\rho] \in H^3(X,\R)$ should be correct, as should the
classical limit of \zhitchin.  In comparing $Z_H$ to $Z_B$ we notice
two points.  First, unlike $Z_B$, $Z_H$ does not depend on a choice of
symplectic coordinates for $H^3(X,\R)$.  Second, $Z_H$ depends on
twice as many degrees of freedom as does $Z_B$ (because $Z_B$ depends
on only half of the coordinates of $H^3(X,\R)$ as explained above.)
So $Z_H$ cannot be equal to $Z_B$.

The situation changes drastically, however, if we combine the B model
with the complex conjugate $\bar{{\rm B}}$ model.  In that case we have
two wavefunctions, $Z_B$ and $\bar{Z_B}$, and we can consider the
product state
\eqn\psidef{
\Psi = Z_B \otimes \bar{Z_B}.
} 
(One could more generally consider a density matrix that is a sum of
such pure product states.)  This product state sits inside a doubled
Hilbert space, obtained from quantization of a phase space which is
also doubled, from $H^3(X,\R)$ to $H^3(X,\C)$.  This doubled phase
space has a polarization which does not depend on any arbitrary
choice: namely, one can divide it into real and imaginary parts, and
it is natural to ask for the representation of $\Psi$ as a function of
the real parts of all the periods, $\Psi(\re X_I, \re F^I)$.  This
gives a function on $H^3(X,\R)$ which does not depend on any choice of
symplectic coordinates.  This is actually a standard construction in
quantum mechanics: the function one obtains expresses the density in
phase-space corresponding to the wavefunction $Z_B$, and is known as
the ``Wigner function'' of $Z_B$.  It is this Wigner function which we
propose to identify with $Z_H([\rho])$.

We can give an explicit formula for the Wigner function if we start
from a particular representation of $Z_B$, namely the one
corresponding to a basis of A and B cycles $\{ A_I, B^I \}$ in
$H_3(X,\Z)$.  (This choice of polarization corresponds to a certain
limit in the space of possible $\Omega_0$; from now on we suppress
$\Omega_0$ in the notation.)  Then $Z_B$ can be written as $Z_B(X_I)$,
a function of the A cycle periods $X_I$, and we denote the B cycle
periods $F^I$.  Writing $P_I = \re X_I$, $Q^I = \re F^I$, the Wigner
function is given by
\eqn\wigner{
\Omega(P_I, Q^I) = \int d\Phi_I \ e^{- Q^I \Phi_I}\ |Z_B(P_I + i \Phi_I)|^2.
}
This can be identified with $Z_H([\rho])$ if we identify $P_I$ and
$Q^I$ as the (real) periods of the class $[\rho] \in H^3(X,\R)$.

At least at string tree level (which in this context means large
$\rho$) we can verify that this identification is correct.  Namely, in
that limit, $Z_B$ is dominated by the tree level free energy $F_0$,
and writing $Z_B = e^{- \ihalf F_0}$, we can make a steepest descent
approximation of the integral over $\Phi$ in \wigner.  The argument of
the exponential is
\eqn\exparg{
- \ihalf F_0(P_I + i \Phi_I) + \ihalf \overline{F_0(P_I + i \Phi_I)} - Q^I \Phi_I.
}
The value of $\Phi$ which extremizes \exparg\ occurs when $Q^I = \re
\partial F_0 / \partial X_I = \re F^I(P + i\Phi)$.  At this $\Phi$,
\exparg\ becomes
\eqn\expargleg{
- \ifour X_I F^I + \ifour \overline{X_I} \overline{F^I} - (\re F^I) (\im X_I) = \ifour X_I \overline{F^I} - \ifour \overline{X_I} F^I.
}
But this is exactly the classical Hitchin action $V_H = -\ifour \int \Omega \wedge \overline{\Omega}$,
evaluated at the value of $\Omega$ for which $\re X_I = P_I$ and $\re F^I = Q^I$.  This establishes
the desired agreement between $Z_H$ and the Wigner function of $Z_B$ at tree level.
In fact, the above relation between the topological string and and $\int \Omega \wedge \overline{\Omega}$
was already noted and used in \OSV, for the purpose of relating the topological string to $4D$ black hole
entropy.  We will discuss this connection in Section 8.

It seems likely that the agreement between $Z_H$ and the Wigner
function will also persist at one loop.  The B model at one loop is
known \BCOV\ to compute the Ray-Singer torsion of $M$, which is a
ratio of determinants of $\bar{\partial}$ operators acting on forms of
various degrees; these determinants should agree with those which
appear in the quadratic expansion of $V_H$ around a critical point.
This basically follows from the fact that the kinetic term is given by
$\int \partial \phi {\overline \partial} \phi$, where $\phi$ is a
$(1,1)$ form and the complex structure is determined by the choice of
critical point.  The difference from the B model is just that here
$\phi$ is not viewed as a chiral field, so we get both the B and
${\bar{\rm B}}$ contributions; the one-loop contribution in the B model alone
is a chiral determinant, the holomorphic square root of the determinant of
the Laplacian.

Finally, we note that, by introducing an extra 3-form field $H$,
we can write the action functional $V_H (\rho)$ in a form
that does not involve square roots, just as we did in \maction\
for the A-model,
\eqn\vhhlong{ S = \int_M \rho \wedge H + \int_M \a \cdot {K_a}^b {K_b}^a
+ \int_M (1 - \a \cdot (\rho \wedge H)) \phi,}
where ${K_a}^b$ is defined in \krhorho.
It is easy to see that integrating out the 3-form $H$ and the
Lagrange multiplier $\phi$
leads to the holomorphic volume action $V_H(\rho)$ of \vrho.
The action \vhhlong\ could be useful for a ``quantum foam''
description of the B model parallel to the one discussed above
for the A model.

\newsec{Reducing Topological M-theory to Form Gravities}

In this section, we want to argue that the various form gravity
theories we reviewed earlier arise naturally on supersymmetric cycles
in topological M-theory.  We will discuss various examples, but the
basic idea is always the same: we consider a ``local model'' of a
complete 7-manifold $X$, obtained as the total space of an
$m$-dimensional vector bundle over an $n$-dimensional supersymmetric
(calibrated) cycle $M \subset X$, such that $m+n = 7$,
\eqn\rnbdle{
\matrix{ \IR^m & \rightarrow & X \cr  & & \downarrow \cr && M }. }
This non-compact local model is intended to capture the dynamics of
the 7-dimensional theory when the supersymmetric cycle $M$ shrinks
inside a global compact $X$.  This idea is natural when one recalls
that the geometry of $X$ in the vicinity of a supersymmetric cycle $M$
is completely dictated by the data on $M$.  Thus the local gravity
modes induce a lower-dimensional gravity theory on $M$.  This is
similar to what is familiar in the context of string theory: near
singularities of Calabi-Yau manifolds one gets an effective
lower-dimensional theory of gravity.

After making an appropriate ansatz,
the three-form $\Phi$ on $X$ induces a collection of $p$-form fields on $M$;
the equations of motion of topological M-theory,
\eqn\trformeom{\eqalign{
& d \Phi = 0, \cr
& d *_{\Phi} \Phi = 0,
}}
lead to equations of motion for the $p$-form fields on $M$.  These
equations of motion can be interpreted as coming from a
topological gravity theory on $M$.

Depending on the dimension $n$ of $M$ and the vector bundle we
choose over it, we will have various ansaetze for $\Phi$, leading
to various gravity theories on $M$. For example, the cases $n=3$
and $n=4$ correspond respectively to associative and
coassociative submanifolds, which are familiar examples of
supersymmetric cycles in manifolds with $G_2$ holonomy. In
these two cases, the corresponding vector bundle over $M$ is
either the spin bundle over $M$ (when $M$ is associative) or the bundle of
self-dual 2-forms over $M$ (when $M$ is coassociative). In order
to obtain the other two gravity theories, namely the cases $n=2$
and $n=6$, one has to assume that the bundle \rnbdle\ splits into
a trivial line bundle over $M$ and a bundle of rank $m-1$.
In this case the holonomy group of $X$ is reduced
to $SU(3)$, so that locally $X$ looks like a direct product,
\eqn\xsplit{ X = \IR \times Y,}
where $Y$ is a Calabi-Yau 3-fold of the form \rnbdle, with
$n+m=6$. Notice that apart from supersymmetric 2-cycles and
6-cycles, Calabi-Yau manifolds also contain supersymmetric
3-cycles and 4-cycles. A priori, the form gravity induced on the
latter may be different from the gravity theory obtained on
associative and coassociative cycles in a 7-manifold with
the full holonomy $G_2$.

In the cases $n=3, m=4$ and $n=4, m=3$ we will be closely following
two constructions of local $G_2$ manifolds given in \refs{\BS,\GPP}
and recently discussed in \CGLP.  These two constructions have some
common features which can be conveniently summarized in advance.  We
let $y^i$ denote a local coordinate on the fiber $\IR^m$, and write $r
= y_i y^i$.  The ansatz is $SO(m)$ invariant, so that $\Phi$ depends
only on $r$ and the coordinates on $M$.  We construct a basis of
$1$-forms in the fiber direction as
\eqn\dydef{ \a^i = D_A y^i = dy^i + (Ay)^i,}
where $A$ is the 1-form induced by a gauge connection on $M$
which acts on the $y^i$ in some representation.

The fact that $\Phi$ is
a stable 3-form means that there exists a 7-dimensional
vielbein $e^i$ such that
$$
\Phi = e^{567} + e^5 \wedge (e^{12} - e^{34})
+ e^6 \wedge (e^{13} - e^{42}) + e^7 \wedge (e^{14} - e^{23}).
$$
In the metric $g$ determined by $\Phi$, the $e^i$ form an orthonormal
basis.  We define a triplet of 2-forms $\Sigma^i$ by the formula
\sviaee\ as in the 2-form gravity:
\eqn\sssee{\eqalign{
\Sigma^1 & = e^{12} - e^{34}, \cr
\Sigma^2 & = e^{13} - e^{42}, \cr
\Sigma^3 & = e^{14} - e^{23}.
}}
Then $\Phi$ is written
\eqn\ansphi{
\Phi = e^{567} + e^i \wedge \Sigma^i.
}

To verify the equations of motion, we
will also need the expression for $*_{\Phi} \Phi$,
derived straightforwardly by expanding in the $e^i$:
\eqn\ansstphi{
*_{\Phi} \Phi = - {1 \over 6} \Sigma_i \wedge \Sigma^i
+ {1 \over 2} \epsilon^{ijk} e^i \wedge e^j \wedge \Sigma^k.
}
In fact, it is convenient to consider a slightly more general form
of $\Phi$:  namely, rescaling
the first three $e^i$ by $f$ and other four by $g$, we obtain
\eqn\ansphigen{
\Phi = f^3 e^{567} + f g^2 e^i \wedge \Sigma^i,
}
and
\eqn\ansstphigen{
*_{\Phi} \Phi = - {1 \over 6} g^4 \Sigma_i \wedge \Sigma^i
+ {1 \over 2} f^2 g^2 \epsilon^{ijk} e^i \wedge e^j \wedge \Sigma^k.
}


\subsec{$3D$ Gravity on Associative Submanifolds}

One local model for a $G_2$ space $X$ is obtained by choosing $X$ to
be the total space of the spin bundle over a 3-manifold $M$.  In this
case, with our ansatz, the field content and equations of motion of
topological M-theory on $X$ reduce to those of Chern-Simons gravity on
$M$; in particular, the condition that $X$ has $G_2$ holonomy implies
that $M$ has constant sectional curvature.

First, let us show that the field content of topological M-theory
on $X$ can be naturally reduced to that of Chern-Simons gravity on
$M$.  This amounts to constructing an ansatz for the associative 3-form $\Phi$ in
terms of forms on $M$.  We write it in the general form \ansphigen\ and
then impose the condition that
$e^1$, $e^2$, $e^3$, $e^4$ are constructed out of an $SU(2)$ connection
on $M$ acting on the spin bundle, as we explained earlier in \dydef:  $e^i =
\a^i$, $i=1, \dots, 4$.  For convenience we also relabel $e^5, e^6,
e^7$ as $e^1, e^2, e^3$, so finally the form of $\Phi$ is
\eqn\ansphixxx{ \Phi = f^3 e^{123} + fg^2 e_i \wedge \Sigma^i, }
where
\eqn\sssee{\eqalign{
\Sigma^1 & = \a^{12} - \a^{34}, \cr
\Sigma^2 & = \a^{13} - \a^{42}, \cr
\Sigma^3 & = \a^{14} - \a^{23}.
}}
Further assume that $f$, $g$ depend only on the radial coordinate $r$,
with $f(0) = g(0) = 1$.  Then along $M$, the only fields (undetermined
functions) in our ansatz are the dreibein $e^i$ and the $SU(2)$
connection $A^i$.  These are exactly the fields of 3-dimensional
gravity in the first-order formalism, and they can be organized
naturally into a complexified gauge field, as we discussed before.

Now we want to check that the equations of motion of topological M-theory
reduce with our ansatz to those of 3-dimensional gravity.  This amounts
to evaluating $d \Phi$ and $d *_\Phi \Phi$ directly, using \ansstphigen.
One finds that if
\eqn\fgansthreed{\eqalign{
& f(r) = \sqrt{3 \Lambda} (1+r)^{1/3}, \cr
& g(r) = 2 (1+r)^{-1/6},
}}
then $d \Phi = 0$ becomes equivalent to
\eqn\dedaeomxxx{\eqalign{
& de = - A \wedge e - e \wedge A, \cr
& dA = - A \wedge A - \Lambda e \wedge e.
}}
The conditions \dedaeomxxx\ are precisely the equations of motion
in 3D Chern-Simons gravity,
\eqn\ccseom{ d \CA + \CA \wedge \CA = 0,}
based on the gauge group $G$ as indicated in Table 1.
Furthermore one can check that $d *_\Phi \Phi = 0$ is automatic 
provided that \fgansthreed--\dedaeomxxx\ are satisfied.  So with this particular
ansatz, the equations of motion of topological M-theory do 
indeed reduce to those of 3-dimensional gravity.


\subsec{$4D$ Gravity on Coassociative Submanifolds}

Another local model of a $G_2$ manifold is obtained by choosing
$X$ to be the bundle of self-dual 2-forms over a 4-manifold $M$.
We will see that in this case the effective gravity theory on $M$
is the 2-form gravity we considered in Section 3.3.

First, let us show that the field content of topological M-theory on
$X$ can be naturally reduced to that of 2-form gravity on $M$.  This
amounts to constructing an ansatz for the associative 3-form $\Phi$ in
terms of forms on $M$.  We write it in the general form \ansphigen\
and then impose the condition that $e^5$, $e^6$, $e^7$ are constructed
out of an $SU(2)$ connection on $M$ acting on the bundle of self-dual
2-forms, as we explained earlier: $e^5 = \a_1$, $e^6 = \a_2$, $e^7 =
\a_3$.  Further assume that $\Sigma^i$ are purely tangent to $M$, and
that $f$, $g$ depend only on the radial coordinate $r$, with $f(0) =
g(0) = 1$.

Thus along $M$ the associative 3-form $\Phi$ can be simply written:
\eqn\phisss{
\Phi = \a^{123} + \a^1 \wedge \Sigma^1 + \a^2 \wedge \Sigma^2 + \a^3 \wedge \Sigma^3.
}
Since we constructed both $\Phi$ and $\Sigma$ from the vielbein, which
determines the metric, the metrics on $M$ which can be reconstructed
from $\Phi$ and $\Sigma$ must agree. It is gratifying that this can be
seen explicitly, as we now do:  recall the expression for the $G_2$ metric
in terms of $\Phi$ from \ggss,
\eqn\ggssx{
\sqrt{g}\,g_{jk} = - {1 \over 144} \Phi_{j i_1 i_2}
\Phi_{k i_3 i_4} \Phi_{i_5 i_6 i_7} \epsilon^{i_1 \ldots i_7}.
}
We wish to consider the components of the metric along $M$, $g_{jk}$,
where $j,k = 1, \ldots, 4$. Normalizing the normal directions to have
length scale $1$, we can view $g$ in \ggssx\ as the determinant
of the 4-dimensional metric on $M^4$.  Also, notice that if $j,k = 1,
\ldots, 4$, then none of the $\Phi$-components can be $e^{567}$.
Hence, all the components of $\Phi$ in \ggssx\ should contain
$\Sigma^i$, {\it cf.} \phisss.

Now, let us consider the combinatorial factors in \ggss.
Since indices $j$ and $k$ are assumed to take values from
$1$ through $4$, one of the indices $i_1$ or $i_2$
in $\Phi_{j i_1 i_2}$ can take values 5, 6, or 7.
Similarly, there are two choices to assign a normal
direction to $i_3$ or $i_4$, and three choices in the last
factor, $\Phi_{i_5 i_6 i_7}$. In total, we get a combinatorial
factor $12=2 \cdot 2 \cdot 3$ and we can write the metric in 
the form
\eqn\gviasssxx{
\sqrt{g}\,g_{ab} = - {1 \over 12} \Sigma^i_{a a_1} \Sigma^j_{b a_2}
\Sigma^k_{a_3 a_4} \epsilon^{ijk} \epsilon^{a_1 a_2 a_3 a_4}.
}
This is exactly the expression \gviasss\ for the 
metric on $M$ in the 2-form gravity.

So we have written $\Phi$ in terms of an $SU(2)$ triplet of two-forms
$\Sigma^i$, which by construction obey the constraint \ssconstr, and
an $SU(2)$ gauge connection, which we used to define the $\a^i$.
These are precisely the fields of the two-form gravity theory we
considered above, which has self-dual Einstein metrics on $M$ as its
classical solutions.  Note that from the viewpoint of the
4-dimensional form theory of gravity, the $SU(2)$ gauge indices
$i=1,2,3$ and the 4-dimensional spacetime indices of $\Sigma^i_{ab}$
are unrelated.  However, we have seen that {\it in the context of
topological M-theory the 3} $SU(2)$ {\it gauge indices are unified
with the 4 spacetime indices to give a 3-form in 7 dimensions.}  This by
itself is rather satisfactory, and suggestive of a deep role for
topological M-theory in the context of 4-dimensional quantum gravity.

Next we want to argue that the field equations of topological M-theory reduce
to those of the two-form gravity theory on $M$.  First consider the 
equation $d \Phi = 0$.  A direct computation shows that with the choice
\eqn\fgansfourd{\eqalign{
& f(r) = (1+r)^{-1/4}, \cr
& g(r) = \sqrt{2\Lambda} (1+r)^{1/4},
}}
the condition $d \Phi = 0$ becomes equivalent to
\eqn\fgcondf{\eqalign{
& D_A \Sigma = 0,\cr
& F \wedge \Sigma = 0.
}}
In fact, the latter equation follows from the former by applying $D_A$ to 
both sides; so we just have to impose $D_A \Sigma = 0$, which means
that $A$ is the $SU(2)_+$ part of the spin connection.  Note that
this also implies that $F$ is self-dual in the metric
induced by $\Sigma$, $F = F_+$; this follows from the fact that $F$
is $SU(2)_+$ valued, and the symmetry $R_{abcd} = R_{cdab}$ of the
Riemann tensor, which is shared by $F$.

Similarly, one finds that $d *_\Phi \Phi = 0$ can be satisfied provided
that
\eqn\fgconds{
F_+ = {\Lambda \over 12} \Sigma.
}
So altogether, the equations of motion of topological M-theory imply
\eqn\fgcondc{\eqalign{
& D_A \Sigma = 0,\cr
& F = {\Lambda \over 12} \Sigma.
}}
These agree precisely with \eomfoursd.
In sum, the field content and equations of motion of the self-dual version 
of two-form gravity agree
with those of topological M-theory, when we make a special ansatz for
$\Phi$.

\subsec{$6D$ Topological Strings}

Finally let us consider the case $n = 6$.  In this case $X$ is a real
line bundle over the 6-dimensional $M$, and we choose it to be trivial
--- either $X = M \times \R$ or its compactification $X=M\times \S^1$.
Let $\R$ be parameterized by $t$.  Then a natural ansatz for $\Phi$ is
\eqn\phisixd{
\Phi = \rho + k \wedge dt,
}
where $\rho$ and $k$ are respectively a 3-form and 2-form on $M$.
If $\Phi$ is a stable 3-form on $X$, then $\rho$ and $k$ are stable on $X$, so
as we discussed earlier, they define respectively an almost complex structure and 
a presymplectic structure on $X$, and
if we impose also the conditions \compata--\compatb\ 
then these two structures are compatible.  In that case they give an $SU(3)$ structure on $X$.
The condition that this $SU(3)$ structure is integrable,
\eqn\sutintegr{\eqalign{
& dk = 0, \cr
& d(\rho + i \hat{\rho}) = 0,
}}
is equivalent to the 7-dimensional equations of motion $d \Phi = 0$, $d *_\Phi \Phi = 0$.
So with this ansatz, topological M-theory on $X$ reduces to a theory on $M$, describing variations
of $k$ and $\rho$, for which the classical solutions are Calabi-Yau 3-folds.

What 6-dimensional theory is this?  As they are usually conceived,
neither the topological A model nor the topological B model alone fits
the bill: at least perturbatively, the A model just describes
variations of $k$, and the B model those of the holomorphic 3-form
$\Omega = \rho + i \hat{\rho}$.  The theory we are getting on $M$ is a
combination of the A and B models --- with a slight coupling between
them, expressed by the constraints \compata--\compatb.  In support
of this point of view, note that after imposing \compata--\compatb\
the action $V_7(\Phi)$ can be simply expressed in terms of $\rho$ and
$k$: it becomes simply
\eqn\splitaction{
V_7 (\Phi) = 3 V_S(k) + 2 V_H(\rho), } where $V_S$ and $V_H$ are the
6-dimensional symplectic and holomorphic volume functionals introduced
in Section 4.3.  As we discussed in Section 5, these functionals
correspond respectively to the A model and the B + $\bar{{\rm B}}$
model.

It is natural to conjecture that this 7-dimensional construction is in
fact related to the nonperturbative completion of the topological
string, which we expect to mix the A and B models, and to related
phenomena such as the topological S-duality conjectured in
\refs{\NV,\NOV} (see also \Kapustin).
While this picture is far from complete, there is one encouraging sign,
which we will describe further in the next section.


\newsec{Canonical Quantization of Topological M-theory and S-Duality}

In the last section we discussed a possible relation between
topological M-theory on $X = M \times \S^1$ and topological string
theory on $M$.  In particular, we found that the classical reduction,
obtained just by considering fields which are independent of the
coordinate along $\S^1$, leads to a combination of two systems, which
are in the universality classes of the topological A and B models.
Recently, there have been some hints that the A and B model could be
coupled to one another.  In this section we discuss how such a
coupling could arise through canonical quantization, and how this
relates to the notion that the topological string partition function
is a wave function.

We begin by considering the 4-form version of topological M-theory.
To perform the canonical quantization of Hitchin's action $V_7(G)$ is
a nontrivial problem, because of the usual subtleties involved in
quantizing a diffeomorphism invariant theory.  Moreover, we should
note that we are viewing Hitchin's action only as an effective action,
which we are using just to extract some basic facts about the Hilbert
space.  For this purpose it is enough to work roughly, although a more
precise treatment would certainly be desirable.

So let us consider the 7-dimensional gravity theory \VG\ on a manifold
$X = M \times \R$, where $M$ is a compact 6-manifold and $\R$ is the
``time'' direction. We decompose the 3-form gauge field $\Gamma$ as
$$
\Gamma = \gamma + \beta \wedge dt,
$$
where $\gamma$ and $\beta$ are a 3-form and 2-form respectively, with
components only along $M$.  Similarly decompose $G$ and $*_G G$ as
\eqn\decomp{\eqalign{
G &= \sigma + \hat\rho \wedge dt, \cr
*_G G &= \rho + k \wedge dt.
}}
Then write
\eqn\eqg{\eqalign{
G &= G_0+d\Gamma \cr
  &= (\sigma_0 + d\gamma) + (\hat\rho_0 + d\beta + \dot{\gamma}) \wedge dt,
}}
so that
\eqn\Gexp{\eqalign{
& \hat{\rho} = \hat{\rho}_0 + d\beta + \dot\gamma, \cr
& \sigma = \sigma_0 + d\gamma.
}}
The configuration space is spanned by the components $(\gamma,\beta)$ of
the gauge field $\Gamma$.  Their conjugate momenta are
\eqn\conjmom{
\pi_\gamma = {7 \over 4} \rho,
\qquad \pi_\beta = 0.}
The longitudinal component $\beta$ is an auxiliary field;
it imposes the constraint 
\eqn\vsevconstr{
{\delta V_7 \over \delta \beta} = d\rho = 0,
}
which generates the spacelike, time-independent gauge
transformations $\gamma \to \gamma + d \lambda.$
Hence the reduced phase space which we obtain from 
canonical quantization of the 4-form theory
is parameterized by $(\gamma,\rho)$, where
\eqn\canquanpha{
\gamma \in \Omega^3(M) /\Omega^3_{exact}(M), \qquad
\rho\in \Omega^3_{closed}(M).
}

Now let us compute the Hamiltonian.  Suppose that we impose the
conditions \compata--\compatb\ (we will comment more on the role of
these constraints later.)  Then it is straightforward to verify that
$\rho$ and $\hat\rho$ are related as in Section 4.3, and $\sigma =
\half k \wedge k$.  The action $V_7$ from \VG\ becomes
\eqn\vonconstr{
V_7 = \int_X dt \left( 2 V_H(\rho) + 3 V_S(\sigma(\gamma)) \right).
}
Using \conjmom\ we can construct the Hamiltonian,
\eqn\hrhosigma{
H = 2 V_H(\rho) + 3 V_S(\sigma(\gamma)) - \dot{\gamma} \wedge \pi_\gamma = {3 \over 2} (2 V_S(\sigma(\gamma)) - V_H(\rho)). }
Although our treatment has been rough, we can gain some confidence from the fact
that the Hamiltonian we ultimately obtained at least gives classical equations
of motion agreeing with the Lagrangian formulation; 
namely, it agrees with \hamilt, which indeed defines a flow
giving $G_2$ holonomy metrics.  A more precise treatment (possibly starting from
a different classically equivalent action) would require a better
understanding of the constraints \compata--\compatb; we believe that 
they will turn out to be the diffeomorphism and Hamiltonian constraints, as
usual in diffeomorphism invariant theories.  Indeed, note that \compatb\
is simply the constraint $H=0$.

As usual in the Hamiltonian formalism, we treat $\rho, \gamma$ as the canonical
variables, where $\rho$ is ``momentum'' and $\gamma$ is ``position.''
{}From \conjmom\ we see that they have canonical commutation relations
\eqn\commu{\{ \delta \gamma , \delta \rho \}= \int_M \delta \gamma \wedge \delta \rho.}
Recalling that $V_H$ and $V_S$ were identified respectively with the B
and A models, we see that the Hamiltonian has split into a ``kinetic
term'' involving the B model and a ``potential term'' involving the A
model.  Despite this splitting the A and B models are not independent;
the fact that $\rho$ and $\gamma$ do not commute at the {\it same
point} of $M$ suggests that, for the quantum Calabi-Yau, the
uncertainty principle would prevent measurements of the complex
structure and \kahler\ structure at a given point from being done
simultaneously.  This is an interesting result which deserves more
scrutiny.

One can also ask about the commutation relations between the zero
modes of the A and B model fields, which might be of more direct
interest, because these zero modes are observables on which the
partition function depends.  Some of these zero modes are already
present in the heuristic construction of the phase space which we gave
above.  For example, one can consider variations of $\gamma$ which are
closed, $d(\delta \gamma)=0$; these induce no variation in the
\kahler\ form, but nevertheless affect the nonperturbative A model
partition function (via the coupling to Lagrangian branes) as we
discussed in Section 3.  These variations of $\gamma$ up to gauge
equivalence parameterize an $H^3(M,\R)$ in the phase space, which via
\commu\ is canonically conjugate to the $H^3(M,\R)$ given by the
cohomology class of $\rho$.  This means that the A model variables and
B model variables mix; the parameter playing the role of the
Lagrangian D-brane tension in the A model is conjugate to the 3-form
of the B model.

There is a dual version of the above discussion: if we had started
from the 3-form version of topological M-theory instead of the 4-form
version, we would have written
\eqn\dualvarmr{
\rho=\rho_0+dB,
}
where $B$ is a 2-form on $M$.  (This field is very closely related to
the field that we denoted as $\phi$ that appeared in the B model
topological string.) The phase space then turns out to be spanned by
$B$ and $\sigma$ with the Poisson bracket given by
\eqn\dualpoiss{
\{ \delta B , \delta \sigma \} =\int_M \delta B \wedge \delta \sigma.
}
This pairing agrees with the one we obtained above, except that it
includes different zero modes: instead of having two copies of
$H^3(M,\R)$ we now have the variations of $B$ up to gauge equivalence
which do not change $\rho$, parameterizing an $H^2(M,\R)$, canonically
conjugate to the $H^4(M,\R)$ given by the cohomology class of
$\sigma$.  Hence the $B$-field which couples to the D1-brane of the B
model is conjugate to the \kahler\ parameter of the A model.

In sum, we seem to be finding that even at the level of the zero
modes, i.e. the observables, there is a sense in which the fields of
the A and B models are conjugate to one another.  It is natural to
suspect that this is related to the conjectured S-duality between the
A and B models, which would be interpreted as position/momentum
exchange or electric/magnetic duality in topological M-theory.  In
particular, the fact that nonperturbative amplitudes of the B model
involve the D1-brane and the $B$-field, and the fact that the
nonperturbative amplitudes of the A model involve Lagrangian D-branes
and the $\gamma$ field, suggest that the full nonperturbative
topological string is a single entity consisting of the A and B models
together.

Clearly these ideas should be developed further, but we feel that
there is a beautiful connection here, between the conjectured
S-duality between the A and B models and the fact that topological
M-theory treats their degrees of freedom as conjugate variables.

One might ask how this Hamiltonian quantization is related to the fact
that the B model partition function is a wavefunction, reviewed in
Section 5.2, which was one of our original motivations for introducing
a 7-dimensional topological M-theory.  In the zero mode sector we have
found two conjugate copies of $H^3(X,\R)$, which would be sufficient
to account for both the phase spaces underlying the B model partition
function and the $\bar{{\rm B}}$ model partition function.  This is
reminiscent of Section 5.2 where we saw that we could interpret the
Wigner function as a wavefunction for the combined B and $\bar{{\rm
B}}$ models, with the zero mode phase space $H^3(X, \C)$,
parameterized by the conjugate variables $\re \Omega = \rho$ and $\im
\Omega = \hat \rho$.  On the other hand, as dicussed above, in the
7-dimensional theory $\gamma$ is conjugate to $\rho$.  We are thus
naturally led to identify $\hat \rho = \gamma$.  
This identification was in a sense
predicted by the topological S-duality conjecture, since it
says precisely that the Lagrangian D-branes of the A model are coupled
to the imaginary part of $\Omega$.
It indeed follows
semiclassically if we identify the wavefunction as being given by the
Hitchin functional, $\Psi(\rho) \sim \exp V_H(\rho)$; then we get the
necessary relation 
\eqn\relationwavefn{
\gamma \vert \Psi \rangle = \hat\rho(\rho) \vert \Psi \rangle
}
using $\delta V_H/\delta \rho=\hat \rho$.  This
relation between the potential and the wavefunction is not
unexpected, since the function $V_H$ is quadratic in $\rho$.

Note that we expect to recover the Wigner function \wigner\ by
considering topological M-theory on a 7-manifold $X = M \times \S^1$,
where $M$ is a Calabi-Yau space.  This is natural to expect from
dimensional reduction, making contact with the discussion of Section
5.2.

\newsec{Form Theories of Gravity and the Black Hole Attractor Mechanism}

In the previous sections we have discussed various theories of gravity
in which one reconstructs geometric structures from $p$-forms on the
spacetime $M$.  Although this might seem like an unusual way to get
these structures, a similar phenomenon occurs in superstring theory
compactified on $M$: given a black hole charge, which can be
represented as an integral cohomology class on $M$, the attractor
mechanism fixes certain metric data $g_{\mu\nu}$ of $M$ at the black
hole horizon \refs{\attrone,\attrtwo,\Kallosha,\Kalloshb}.
In other words, it provides a
map\foot{More precisely, it fixes some of the components of $g$; not
all of the moduli are fixed by the attractor mechanism.}
\eqn\attrflow{
Q \mapsto g_{\mu\nu}.
}

In this section, we will discuss a relation between black holes and
Hitchin's functionals.  In particular, we argue that these functionals
also lead to the map \attrflow.  In a sense, the metric flow of the
internal manifold from spatial infinity to the black hole horizon can
be viewed as a geodesic flow with respect to Hitchin's action.  In
fact, Hitchin's picture is more general: {\it it does not assume the
metric to be of the Calabi-Yau form, but derives that from the same
action principle which leads to the relation between the charge and
metric.  The usual attractor mechanism only discusses the zero mode
sector of the metric, whereas Hitchin's action also deals with the
massive modes}.

This link between form theories of gravity and BPS black holes can,
in fact, lead to a fundamental nonperturbative
definition of the gravitational form theory as counting black hole degeneracies
with a fixed charge, as in the recent work \refs{\OSV,\VafaYM}.
This interpretation of the gravitational form theories 
also ``explains'' why one fixes the cohomology class of the form and 
integrates only over massive modes; this corresponds simply to 
fixing the black hole charge.  At least in the cases of $4D$ and $5D$
BPS black holes, we will show that this interpretation is correct
at leading order in the black hole charge; this amounts to the 
statement that {\it the value of the extremized classical action agrees with the
semiclassical black hole entropy}.

\subsec{BPS Black Holes in 4 Dimensions}

We begin with the case of $4D$ BPS black holes in Type IIB string
theory compactified on a Calabi-Yau 3-fold $M$.  In Section 4.3 we
defined Hitchin's ``holomorphic volume'' \defVrho, a functional of a
3-form $\rho$ in six dimensions:
\eqn\holvolsix{
V_H(\rho) = {1 \over 2} \int_M \hat\rho \wedge \rho = - {i \over 4} \int_M \Omega \wedge \bar{\Omega}.
}
Furthermore we noted that, if we hold the cohomology class $[\rho]$
fixed (writing $\rho = \rho_0 + d \beta$), the critical points of
$V_H(\rho)$ yield holomorphic 3-forms on $M$ with real part $\rho$.
So the process of minimizing $V_H$ produces the imaginary part of
$\Omega$ as a function of its real part.  Remarkably, this is exactly
what the attractor mechanism does: fixing the black hole charge $C$
for the theory on $\R^4$, the attractor mechanism produces the value
of $\Omega$ of the Calabi-Yau at the black hole horizon, and the real
part of $\Omega$ is equal to $C^*$, the Poincare dual of $C$.
Therefore it is natural to identify
\eqn\bhcharge{
[\rho] = C^*.
}
Note that the quantization of $C$ matches the fact that $\rho$ is naturally quantized,
if we view it as the field strength of the $2$-form potential $\beta$.
So the holomorphic volume functional $V_H$ is related to the attractor
mechanism at least classically.

Furthermore, the classical value of the action also has a natural
physical meaning: namely, after fixing $C$, the value of $\int \Omega \wedge \bar{\Omega}$ 
at the critical point gives the
leading-order contribution to the black hole entropy at large $C$.
Now consider the {\it quantum} theory with action $V_H$.  The path
integral formally defines a partition function $Z_H(C)$ depending on
the charge,
\eqn\zhol{
Z_H(C) = \int_{[\rho] = C^*} D\rho \exp(V_H(\rho)).  } 
We conjecture
that this path integral computes the exact number of states of the
black hole (or more precisely the index $Z_{{\rm BH}} (C)$ defined in
\OSV, which counts the states with signs):
\eqn\bhcounting{
Z_{{\rm BH}} (C) = Z_H (C).  } 
The main evidence for this conjecture is that if the path integral
\zhol\ exists, it would be a function of $C$ whose leading asymptotics
agree with the black hole entropy --- it would be remarkable if there
were two such functions with natural physical definitions and they
were not equal.  Conversely, one could {\it define} the
nonperturbative quantum theory by the black hole entropy.

Additional evidence for the conjecture \bhcounting\ comes by noticing
that it is essentially the conjecture of \OSV, which identified
$Z_{BH}(C)$ with a Wigner function constructed from the B model
partition function $Z_B$.  Namely, choose a splitting of $H^3(M,\Z)$
into A and B cycles.  Then splitting $C$ into electric and magnetic
charges, $C = (P,Q)$, one has \OSV
\eqn\osvconj{
Z_{{\rm BH}} (C) = \int d \Phi\, e^{i Q^I \Phi_I} |Z_{B}(P + i \Phi)|^2.
}
On the other hand, as we already discussed in Section 5.2, there is
indeed a relation \wigner\ between the B model and Hitchin's theory,
\eqn\osvconjx{
Z_H (C) = \int d \Phi\, e^{i Q^I \Phi_I} |Z_{B}(P + i \Phi)|^2.
}
Recall that Hitchin's theory based on $V_H$ is related not to the B
model but to the B plus ${\bar{\rm B}}$ model; this agrees well with the
fact that this B plus ${\bar{\rm B}}$ model also appears in the counting of
black hole entropy.  This makes one more confident that the connection
between Hitchin's theory and the black hole is direct and deep.

\subsec{BPS Black Holes in 5 Dimensions}

So far we have discussed a relation between $V_H$ and counting of
4-dimensional BPS black hole states obtained from Type II string
theory on $M$.  But as described in Section 4.3, there is also the
functional $V_S$ which makes sense on the 6-manifold $M$; one could
ask whether it is also related to black hole entropy.  In this section
we will argue that it is, and the black holes in question are the ones
in the 5-dimensional theory obtained by compactifying M-theory on $M$.
These BPS black holes can be constructed by wrapping M2-branes over
2-cycles of $M$, and are characterized by a charge $Q\in
H_2(M,\Z)=H^4(M,\Z)$ and a spin $j$.  At first let us set $j=0$.  To
connect the black hole counting to Hitchin's theory based on $V_S$, we
identify
\eqn\qsig{
Q^*=[\sigma].
}  
The attractor value of the moduli in this case is given \refs{\Kalloshb,\bhfive} by
a \kahler\ form $k$, such that $\half k^2=\sigma$; with this value of $k$, the volume of the
Calabi-Yau is proportional to the entropy of the black hole,
\eqn\sbhkah{
S_{BH} \sim \int_M k^3 = \int_M \sigma^{3/2}.
}
In other words, the black hole entropy is given by the classical value of 
$V_S(\sigma)$.
This is automatically consistent with the fact that the black hole
entropy in five dimensions scales as $Q^{3/2}$.  So, in parallel with what 
we did for $V_H$, we conjecture that the partition function $Z_S([\sigma])$ 
of the theory based on $V_S$ counts BPS states of 5-dimensional black holes.

It is possible to extend the foregoing discussion to spinning black holes,
by introducing an additional 6-form field $J$ in the Hitchin action $V_S$.
We denote the integral cohomology class of $J$ by
$j = [J] \in H^6(M,\Z)$; this $j$ can be 
naturally identified with the spin of the black hole.  We consider the action
\eqn\hitj{
V_S(\sigma,J) = \int \sqrt{\sigma^3-J^2},
}
where
$$\sigma^3-J^2=(\sigma_{i_1i_2i_3i_4}\sigma_{j_1j_2j_3j_4}\sigma_{k_1k_2k_3k_4} -J_{i_1i_2j_1j_2k_1k_2}
J_{i_3i_4j_3j_4k_3k_4})\epsilon^{i_1i_2j_1j_2k_1k_2}\epsilon^{i_3i_4j_3j_4k_3k_4}.
$$
It is easy to see that this modification does not change the attractor value
of the \kahler\ form $k$, but changes the classical value of the action to $\sqrt{Q^3-j^2}$,
which agrees with the entropy of the spinning black hole.

We have just argued that the quantum theory based on the extended functional
\hitj\ should count the degeneracies of BPS black holes in five dimensions.
On the other hand, since the perturbative A model counts
exactly these degeneracies \GVtwo, one might expect a direct relation
between the A model and \hitj.  At least for $j=0$, we have already encountered
this relation in Section 5.1, where the quantum foam description of the
A model was related to a Polyakov version of $V_S$.

\subsec{Other Cases}

It is natural to conjecture that the relation between BPS objects and
form gravity theories goes beyond the examples discussed above.  In
particular, it would be interesting to develop this story for the case
of $G_2$ manifolds.  For example, in M-theory compactified on a $G_2$
manifold, we can consider BPS domain walls formed from M5-branes
wrapped on associative 3-cycles.  It is natural to conjecture that the
quantum version of Hitchin's 4-form theory is computing the
degeneracies of these domain walls.  In the Type IIB superstring
compactified on a $G_2$ manifold, one can similarly ask about the
degeneracy of BPS strings obtained by wrapping D5-branes on
coassociative 4-cycles; one might expect a relation
between this counting and the quantum version of Hitchin's 3-form
theory.

\newsec{Topological $G_2$, Twistors, Holography, and $4D$ Gauge Theories}

\noindent In this section, we discuss possible dualities relating three
different theories:

\item{$i)$} gauge theory on a Riemannian 4-manifold $M$;

\item{$ii)$} topological A model on the twistor space, $T(M)$, of
a 4-manifold $M$;

\item{$iii)$} topological M-theory on a 7-manifold $X$,
\eqn\rtribdle{\matrix{ \R^3 & \rightarrow & X \cr  & & \downarrow
\cr && M }}

\noindent As we reviewed earlier, the 7-manifold $X$ admits a
natural metric with $G_2$ holonomy if $M$ is a self-dual (i.e.,
with self-dual Weyl tensor) Einstein 4-manifold.\foot{Such
manifolds are also known as quaternionic K\"ahler manifolds of
dimension 1.} In that case, the $\R^3$ bundle \rtribdle\ is the
bundle of self-dual 2-forms on $M$. Let us compare this to the
corresponding geometric structure on the twistor space $T(M)$.

First, let us recall the definition of the twistor space $T(M)$.
Consider the space of self-dual 2-forms of norm 1.  For each point
on $M$ this gives rise to a 2-sphere. The total space is the
twistor space, $T(M)$, which has a canonical almost complex
structure and also a canonical map to $M$, with fiber being the
twistor sphere ${\bf CP}^1$.  There is a remarkable connection
between self-dual metrics on $M$ (not necessarily Einstein) and
the integrability of the almost complex structure on $T(M)$:
$T(M)$ {\it has an integrable complex structure if and only if}
$M$ {\it is self-dual} \refs{\AHS,\Penrose}. Moreover, $T(M)$ admits a K\"ahler
structure if and only if $M$ is Einstein \Hittwistor\ (see also \Lebrun).
These are the necessary conditions for the existence
of topological A and B models on $T(M)$.

In order to complete this to a string theory we also need
conformal invariance, which is usually guaranteed by a Ricci
flatness condition. This is not the case, however, for the twistor
space $T(M)$, which is not Ricci-flat.  One can complete
$T(M)$ to a Ricci-flat supermanifold by including extra fermionic
directions \Wittentwistor.  We want to explore another way 
of obtaining Ricci-flatness.  As discussed above, the bundle
$X$ of self-dual 2-forms over $M$ has a natural $G_2$ holonomy 
metric, so in particular it is Ricci-flat.  On the other hand,
the boundary of $X$ is precisely the
twistor space,
\eqn\tmbdry{ T(M) = \p X. }
In this sense we could view $X$ as obtained by adjoining a 
radial direction to $T(M)$.

So we could define a topological string theory on the twistor
space $T(M)$ as a holographic dual to topological M-theory on
$X$.  We note that the A model can be
defined on 3-folds which are not necessarily Calabi-Yau;
it has been studied in the mathematical literature on
Gromov-Witten theory \refs{\Giverntal,\KManin,\BManin,\Eguchietal}.
Conversely, using the Gromov-Witten theory on $T(M)$ we can
define topological M-theory on $X$, at least perturbatively.

This holographic duality is reminiscent of our original motivation to look
for a 7-dimensional theory, which would naturally explain the
observation that topological string partition function should be
viewed as a wavefunction.  We also note that, in the present case, the boundary
6-manifold is not stationary under Hitchin's Hamiltonian flow equations; this
reflects the fact that $T(M)$ is not a Calabi-Yau.

\bigskip
\noindent {\it Large $N$ Holography and Gravitational Holography}
\medskip

We are familiar with examples of holography in the context
of open-closed string dualities, where in the large $N$ limit
D-branes wrapping some cycles disappear and the theory
is best described by a new geometry obtained by deleting
the locus of the D-branes, replaced by suitable fluxes.
Via this holographic duality, the open string gauge theory
provides an answer to questions of gravity in a geometry
obtained by a large $N$ transition.

Another kind of duality --- which is somewhat similar to holography ---
is a duality between M-theory on an interval and the heterotic
$E_8\times E_8$ theory living on the boundary \HW.
In this duality, the coupling constant of the heterotic
string controls the size of the interval.
Even though the heterotic string ``lives'' on the boundary,
it can be used, at least in principle,
to study gravitational physics in the bulk.

The relation between the 7-dimensional topological
M-theory on $X$ and the topological string on
a 6-manifold $T(M)$ is more similar to the heterotic/M-theory duality.
In this sense, when we say that the partition function
of a topological string theory on the twistor space
can be regarded as a wave function in a 7-dimensional theory on $X$,
what we mean is a ``gravity/gravity holography.''

Having said that, it is natural to ask:  is there
an open-closed string holographic duality in the present context?
Given that we do not yet have a deep understanding of topological
M-theory, we will limit ourselves to some string-motivated speculations below.

In order to have an open/closed duality, we need to be working in some context where
D-branes exist.  From the point of view of embedding
of the $G_2$ theory in the physical M-theory, it is natural to compactify
on one more circle and obtain a Type IIA string theory compactification
on a $G_2$ manifold.
So let us consider Type IIA on a non-compact $G_2$ manifold $X$ of
the form \rtribdle, with $N$ D-branes wrapped
over the coassociative 4-manifold $M$ (in the full superstring theory
these could be viewed for example as spacetime-filling D6-branes).
By analogy with geometric transitions in Calabi-Yau
3-folds \GopakumarV,
in the large $N$ limit we expect a transition to a new
geometry which can be obtained by removing the locus
of the D-branes,
$$
X \setminus M.
$$
This space is a real line bundle over $T(M)$.
{}From the discussion earlier in this section,
we expect topological M-theory on this 7-manifold
to be related to topological string theory on $T(M)$.
This leads to a natural conjecture that
topological gauge theory on a 4-manifold $M$
is a holographic dual to topological string theory
on the twistor space $T(M)$.  This dovetails in a natural
way with the idea that the topological string partition function
should be viewed as a wave function on the boundary of the 7-dimensional
manifold with $G_2$ holonomy.

What kind of topological gauge theory in four dimensions
should we expect? The most natural conjecture is that
it is the self-dual Yang-Mills theory, which is
related to the D-brane theory for $\CN=2$ strings.
In other words, one might conjecture that the self-dual
Yang-Mills on $M$ is dual to topological strings on $T(M)$,
so that the K\"ahler class of ${\bf CP}^1 \in T(M)$
is identified with the `t Hooft parameter of the dual gauge theory, $t=Ng_s$.
Below, we consider this duality in more detail
for $M=\S^4$ and $T(M)={\bf CP}^3$.

Topological string theory on $T(M)={\bf CP}^3$ is rather trivial
due to the $U(1)$ charge conservation on the worldsheet.
In particular, the free energy is simply given by the cubic classical
triple intersection of ${\bf CP}^3$.
This agrees with the fact that self-dual Yang-Mills
is also trivial in perturbation theory.
However, it is known that topological strings can be made more
interesting by turning on higher charge $(q,q)$ form operators
with $q=2,3$. The most natural one is the volume form with $q=3$,
which preserves all the symmetries of ${\bf CP}^3$.
Once we add an operator $s \Phi_{3,3}$,
the topological A model string on ${\bf CP}^3$
becomes non-trivial and receives all order corrections.
Thus, the partition function of the perturbed A model
is a function of two independent variables,
\eqn\zztop{ Z^{top} (g_s, s^2 e^{-t} ). }
The fact that the combination $s^2 e^{-t}$ appears
follows from charge conservation of the topological A model.

One possibility is to look for a deformation of the self-dual
Yang-Mills corresponding to the deformation of
the topological A model by the operator $s \Phi_{3,3}$.
In a realization {\it \'{a} la} Siegel \refs{\Siegel,\CSiegel},
\eqn\fgaction{ S=\int d^4x\,\Tr\,F \wedge G, }
the self-dual Yang-Mills is written in terms of
a $U(N)$ adjoint valued self-dual 2-form $G$
and the curvature of a $U(N)$ connection, $F$.
We can deform the action \fgaction\ by the term
$\epsilon G \wedge G$, which (perturbatively)
leads to the full Yang-Mills theory.
It is natural to ask whether this deformation is dual
to the deformation of the A model on ${\bf CP}^3$ by $s \Phi_{3,3}$.
Notice, that bosonic Yang-Mills on $\S^4$ has partition
function which depends on the radius of the 4-sphere, $R$,
the coupling constant of Yang-Mills theory, $g_{YM}^2$,
and the rank $N$ of the gauge group.
Due to the running of the coupling constant
only one combination of $g_{YM}^2$ and $R$ appears.
It is not unreasonable to suppose that with a suitable
choice of the parameter map (which should involve
some kind of Fourier transform) we have
\eqn\zzconject{
Z^{top}_{{\bf CP}^3}(g_s,s^2 e^{-t})\leftrightarrow Z^{YM}_{\S^4}(g_{YM}^2,N) }
It would be very interesting to further develop and check this
conjecture.  If correct, it would allow one to place the appearance of
the higher-dimensional twistor space $T(M)$ in the large $N$ limit of
gauge theory on $M$ into the context of more familiar large $N$
dualities, {\it e.g.} the duality between Chern-Simons theory on $S^3$
and topological strings on the 6-dimensional resolved conifold
\GopakumarV.

\newsec{Directions for Future Research}

In this paper we have discussed the fact that many theories of gravity
fall into the general class of ``form gravity theories,'' and that
they seem to be unified into a 7-dimensional theory of gravity,
topological M-theory, related to $G_2$ holonomy metrics.  We have seen
in particular that this 7-dimensional theory contains the A and B
model topological strings, which appear as conjugate degrees of
freedom.  We have also seen connections with 3-dimensional
Chern-Simons gravity and a 4-dimensional form theory of gravity ---
the topological sector of loop quantum gravity.

Intriguing as this list is, we view this as only a modest beginning:
the connections we have outlined raise many new questions which need
to be answered.  In order to understand better the non-perturbative
aspects of the A and B models, and particularly their implementation
in the context of topological M-theory, we need to understand better
the relation between these models and M-theory.  In particular, it
seems natural to try to explain the S-duality relating the A and B
models using the S-duality of Type IIB superstrings.  This could be
embedded into the present discussion if we include one more dimension
and consider 8-dimensional manifolds of special holonomy.  The natural
candidate in that dimension are manifolds with $Spin(7)$ holonomy.  It
seems that we also need to include this theory in our discussion of
dualities to get a better handle on the S-duality of the A/B models.

Another natural question we have raised relates to the interpretation
of the topological M-theory: does it indeed count domain walls?  This
is a very natural conjecture based on the links we found between form
theories of gravity and the counting of black hole states.  It would
be important to develop this idea more thoroughly.

Another question raised by our work is whether one can reformulate the
full M-theory in terms of form theories of gravity.  This may not be
as implausible as it may sound at first sight.  For example, we do
know that ${\CN=2}$ supergravity in 4 dimensions, which is a low
energy limit of superstrings compactified on Calabi-Yau manifolds, has
a simple low energy action: it is simply the covariantized volume form
on $(4|4)$ chiral superspace \Sokatchev.  In fact, more is true: we could include
the Calabi-Yau internal space as and write the leading term in the
effective action as the volume element in dimension $(10|4)$.  The
internal volume theory in this case would coincide with that of
Hitchin.  Indeed, this is related to the fact that topological string
amplitudes compute F-terms in the corresponding supergravity theory.
Given this link it is natural to speculate that the full M-theory does
admit such a low energy formulation, which could be a basis of another
way to quantize M-theory --- rather in tune with the notion of quantum
gravitational foam.

We have also discussed a speculation, motivated by topological
M-theory, relating gauge theories on $M^4$ to topological strings on
its twistor space.  This connection, even though it needs to be stated
more sharply, is rather gratifying, because it would give a
holographic explanation of the fact that in the twistor correspondence
a 4-dimensional theory gets related to a theory in higher dimensions.
It would be very interesting to develop this conjectural relation; the
potential rewards are clearly great, as a full understanding of the
duality could lead to a large $N$ solution of non-supersymmetric
Yang-Mills.

\bigskip
\bigskip

{\bf{Acknowledgements:}} We are grateful to M.~Atiyah, J.~de Boer, R.~Bryant, C.~LeBrun,
J.~Louis, H.~Ooguri, M.~Ro\v{c}ek, L.~Smolin, C.~Taubes, E.~Verlinde, and S.-T.~Yau
for valuable discussions. We would like to thank the 2004 Simons
Workshop on Mathematics and Physics and the Aspen Workshop ``Strings,
Branes and Superpotentials,'' which led to the development of
many of the ideas in this paper.  We also thank the organizers of the
Strings '04 conference in Paris for providing a stimulating
environment where part of this work was done. S.G. and A.N. would
like to thank Caltech Particle Theory Group, where part of this work
was done, for kind hospitality. This work was conducted during the
period S.G. served as a Clay Mathematics Institute Long-Term Prize
Fellow. S.G. was also supported in part by RFBR grant 04-02-16880.
The research of A.N. and C.V. was supported in part by NSF grants
PHY-0244821 and DMS-0244464. The research of R.D. was partly supported
by FOM and the NWO Spinoza premium.

\appendix{A}{Hitchin's Hamiltonian Flow and Geometry of $\CN=1$ String Vacua}

The geometric structures which appear in the 7-dimensional
topological gravity are reminiscent of the geometries that arise
in $\CN=1$ superstring compactifications. For example, 7-manifolds
with $G_2$ holonomy are classical solutions in 7-dimensional
topological gravity and, on the other hand, are $\CN=1$ vacua of M-theory.
This relation can be extended to
6-manifolds with $SU(3)$ structure which play an important role
in understanding the space of string vacua with minimal ($\CN=1$)
supersymmetry, and which we briefly review in this appendix;
see \refs{\Louis,\Cardoso,\Kachrunew,\Gurrieri,\Kaste,\Becker,\Gauntletttors,\Grana,\Gurrierii} for more details.

Let $M$ be a 6-manifold with $SU(3)$ structure.  Such $M$ are
characterized by the existence of a globally defined, $SU(3)$
invariant spinor $\xi$, which is the analog of the covariantly
constant spinor one has on a Calabi-Yau manifold. In general, instead of
$\nabla \xi = 0$ we have
\eqn\tcovd{ \nabla^{(T)} \xi = 0, }
where $\nabla^{(T)}$ is a connection twisted by torsion $T$.
Roughly speaking, the intrinsic torsion $T$ represents the
deviation from the Calabi-Yau condition. Its $SU(3)$
representation content involves five classes, usually denoted $\CW_i$
\refs{\Salamon, \CSalamon}:
\eqn\tcomps{ T \in \CW_1 \oplus \CW_2 \oplus \CW_3 \oplus \CW_4
\oplus \CW_5. }
In order to describe the geometric meaning of each of these
components, it is convenient to introduce a 2-form $k$ and a
3-form $\Omega$,
\eqn\jomviaxi{\eqalign{
& k = -i \xi^{\dagger} \Gamma_{mn} \Gamma_7 \xi, \cr
& \Omega = -i \xi^{\dagger} \Gamma_{mnp} (1+\Gamma_7)\xi, }}
which satisfy
\eqn\jomzero{ k \wedge \Omega = 0. }
On a Calabi-Yau manifold, the 2-form $k$ would be the usual
K\"ahler form, while $\Omega$ would be the holomorphic volume form.
In particular, $M$ is a Calabi-Yau manifold if and only if $dk = 0$, $d
\Omega = 0$. On a general manifold $M$ with $SU(3)$ structure,
these equations are modified by the components of torsion,
\eqn\djdom{\eqalign{
& dk = - {3 \over 2} \i (W_1 \bar \Omega) + W_4 \wedge k + W_3, \cr
& d \Omega = W_1 k^2 + W_2 \wedge k + \bar W_5 \wedge \Omega,
}}
where
%
\eqn\wwwww{\eqalign{
& W_1 \in \Omega^0 (M), \cr
& W_2 \in \Omega^2 (M), \cr
& W_3 = \bar W_3 \in \Omega_{{\rm prim}}^{2,1} (M) \oplus
\Omega_{{\rm prim}}^{1,2} (M),\cr
& W_4 = \bar W_4 \in \Omega^1 (M), \cr
& W_5 \in \Omega^{1,0} (M).
}}

A particularly interesting class of manifolds with $SU(3)$
structure are the so-called half-flat manifolds. In superstring
theory, they play an important role in constructing realistic
vacua with minimal ($\CN=1$) supersymmetry, and can be viewed as
mirrors of Calabi-Yau manifolds with (a particular kind of) NS-NS
fluxes \Louis. Since under mirror symmetry 3-forms are mapped into
forms of even degree, on half-flat manifolds one might expect
``NS-NS fluxes'' represented by forms of even degree \Vafa. In fact,
as we explain in a moment, on a half-flat manifold $M$ we have
$$
d (\i \Omega) \sim F_4^{NS}.
$$

Half-flat manifolds are defined by requiring certain torsion
components to vanish,
\eqn\wwzero{ \r \CW_1 = \r \CW_2^- = \CW_4 = \CW_5 = 0. }
It is easy to see from \djdom\ that this is equivalent to the
conditions
\eqn\djdomhalff{\eqalign{
& d (k \wedge k) = 0, \cr
& d (\r \Omega) = 0.
}}
If as usual we define $\sigma = \half k \wedge k$ and $\Omega =
\rho + i \hat \rho$, we can write these equations in the familiar
form
\eqn\sutformrels{\eqalign{ & d \rho =0, \cr & d \sigma =0, }}
with an additional constraint $\rho \wedge k = 0$. This is
precisely the structure induced on a generic 6-dimensional
hypersurface inside a $G_2$ manifold, where $\rho$ is the
pull-back of the associative 3-form $\Phi$ and $\sigma$ is the
pull-back of the coassociative 4-form $* \Phi$.  In particular,
using Hitchin's Hamiltonian flow which we reviewed in Section 4.5,
a half-flat $SU(3)$ structure on $M$
can always be thickened into a $G_2$ holonomy
metric on $X = M \times (a,b)$.

So the phase space underlying Hitchin's Hamiltonian flow consists precisely of 
the half-flat manifolds which appear in $\CN=1$ string compactifications with
fluxes and/or torsion,
\eqn\hhitchin{ \CP_{{\rm Hitchin}} = \{ M^6_{{\rm half-flat}} \}. }
Moreover, the ground states are related to stationary
solutions of Hitchin's flow equations, namely Calabi-Yau
manifolds,
\eqn\cygrdstate{ \vert {\rm vac} \rangle \quad \quad
\Leftrightarrow \quad \quad M^6 = {\rm Calabi-Yau}. }

It is tempting to speculate that all $\CN=1$ string vacua can be
realized in topological M-theory.

\listrefs
\end